\documentclass{nature}
\usepackage[utf8]{inputenc}

\title{Narrowing the gap on heritability of common disease by direct estimation in case-control GWAS}

\author{David Golan$^1$ \& Saharon Rosset$^1$}

\date{April 2013}

\usepackage{graphicx}
\usepackage{float}

\usepackage{color}
\usepackage{ulem}

\usepackage[T1]{fontenc}
\usepackage{amsmath}
\usepackage{amssymb}
\usepackage{esint}
\newcommand{\srdel}[1]{}
\newcommand{\srdelmath}[1]{\textcolor{blue} {\text{~\\MATH DELETED HERE\\\\}}}
\newcommand{\srdelfig}[1]{\textcolor{blue} {\text{~\\FIGURE(s) DELETED HERE\\\\}}}

\newcommand{\citep}{\cite}
\begin{document}

\maketitle

\begin{affiliations}
 \item Department of Statistics and Operations Research, Tel-Aviv University.
 \end{affiliations}

\section*{Summary paragraph}

One of the major developments in recent years in the search for missing heritability of human phenotypes is the adoption of linear mixed-effects models (LMMs) to estimate heritability due to genetic variants which are not significantly associated with the phenotype\citep{yang2010common}. 
A variant of the LMM approach has been adapted to case-control studies and applied to many major diseases\citep{Lee2011estimating,do2011web,lee2012estimating,lee2013estimation}, successfully accounting for a considerable portion of the missing heritability. For example, for Crohn's disease their estimated heritability was 22\% compared to 50-60\% from family studies.
In this letter we propose to estimate heritability of disease directly by regression of phenotype similarities on genotype correlations, corrected to account for ascertainment. We refer to this method as genetic correlation regression (GCR). Using GCR we estimate the heritability of Crohn's disease at 34\% using the same data.
We demonstrate through extensive simulation that our method yields unbiased heritability estimates, which are consistently higher than LMM estimates.
Moreover, we develop a heuristic correction to LMM estimates, which can be applied to published LMM results. Applying our heuristic correction increases the estimated heritability of multiple sclerosis from 30\%\citep{lee2013estimation} to 52.6\%.

\section*{Main text}
The mystery of the “missing heritability” is a term commonly used to denote the gap between the expected heritability of many common diseases, as estimated by family and twin studies, and the overall additive (narrow-sense) heritability obtained by accumulating the effects of all single-nucleotide polymorphisms (SNPs) that have been found to be significantly associated with these conditions in genome-wide association studies (GWASs)\citep{bloom2012finding,eichler2010missing,maher2008case,manolio2009finding}. Many diseases which comprise a considerable portion of the health-care burden display such a gap, including type-1 and type-2 diabetes, bipolar disorder, schizophrenia, Alzheimer's disease, multiple sclerosis and Parkinson's disease.


Researchers have proposed several hypothetical solutions to this mystery. These theories include rare causative variants, which are undetected by the current GWAS methodology, common variants with small effects, which do not pass the significance threshold and are therefore unaccounted for, gene-gene and gene-environment interactions which are overlooked by the additive model assumed by the GWAS scheme, epigenetic effects and more\citep{eichler2010missing,manolio2009finding,zuk2012mystery}.


Clearly, different theories have major implications for our understanding of human disease, and also dictate different strategies for discovery of the 
underlying genetic causes of disease. For example, identifying rare variants requires a focus on deep sequencing\citep{nielsen2011genotype}, while detecting small effects of common variants requires increasing the sample sizes dramatically, or conducting massive meta-analyses\citep{visscher2008sizing}. Effective planning of genetic research can therefore only be guided by a satisfactory and well founded allocation of missing heritability to its various possible sources and causes. 
Additionally, accurate estimates of heritability can facilitate better personalized genetic risk predictions\citep{dudbridge2013power}.


Yang et al.\citep{yang2010common} pioneered the use of LMMs to estimate heritability of continuous traits (e.g. height) from GWAS data, while accounting both for significantly and insignificantly associated SNPs, thus providing an estimate of the total heritability explained by common SNPs. Their method was applied to numerous phenotypes including human height, body mass index, von Willebrand factor\citep{yang2011genome}, gene expression\citep{price2011single} and intelligence\citep{visscher2008heritability}. The LMM method was later adapted to dichotomous disease phenotypes by Lee et al.\citep{Lee2011estimating}, assuming the well-known liability threshold model\citep{dempster1950heritability}. They showed that for the case of a random (unascertained) population sample the LMM method can be used to estimate heritability by 
applying it directly to the observed phenotype as if it were continuous, and correcting the resulting "observed scale" estimate as suggested by Dempster and Lerner\citep{dempster1950heritability}. 

The ascertained case-control scenario, which is relevant for most GWAS studies, proved more challenging, as the enrichment of cases violates the assumption of normality of the genetic component critical for LMM estimation. Lee et al. proposed a complex mathematical solution for this case, and demonstrated that it worked in limited simulations. 
They applied their method to three phenotypes from the Wellcome Trust Case-Control Consortium (WTCCC)\citep{WTCCC}: Crohn's disease (CD), bi-polar disorder (BD) and type-1 diabetes (T1D), and it has since been applied to many major diseases including 
schizophrenia\citep{lee2012estimating}, Alzheimer's disease,  multiple sclerosis\citep{lee2013estimation} and Parkinson's disease\citep{do2011web}.
In all these cases, accounting for insignificant SNPs resulted in much higher estimates of heritability than the estimates obtained by accumulating SNPs which were found significant in GWAS.


The basic idea of heritability estimation methods in GWAS is that individuals who are more correlated genetically are more likely to have similar phenotypes. The strength of this connection depends on the heritability -- higher heritability implies stronger connection. The genetic correlation of every pair of individuals can be estimated from their genotypes. 
Since the individuals in the study are unrelated, the correlations between their genotypes are typically small, but non-zero. By accumulating the information across all pairs of individuals in the study, one can leverage these minor differences to separate the phenotypic variance into its genetic and environmental components, resulting in an estimate of heritability.


Ascertained case-control studies pose a considerably harder challenge to deal with theoretically than non-ascertained (prospective or observational) studies, as the fact that cases are over-represented in the study creates a wide range of artifacts.
The normality of the genetic component, assumed by the liability threshold model, is violated by  over-sampling of cases, as is the assumption that the genetic and environmental components are independent. Methods that critically rely on these assumptions, like LMM, can not be applied anymore, without requiring complex and potentially questionable adjustments.

We therefore developed our GCR method for estimating heritability of polygenic phenotypes in ascertained case-control studies. The key novelty of our method is that we model the selection process directly and account for ascertainment by conditioning on the selection of observed individuals. This is contrary to the existing methods of Lee et al.\citep{Lee2011estimating} and Zhou et al.\citep{zhou2013polygenic}, which apply methodologies suited for prospective studies of a continuous normal phenotype, and then attempt to correct the estimates so that they account for ascertainment.
 
Technically, we adopt the commonly used liability threshold model\citep{Lee2011estimating,dempster1950heritability}, and derive analytically the relationship between the genetic correlation of any two individuals, and the similarity between their phenotypes, conditional on the fact that both individuals were selected for the study. While this relationship has no closed-form expression, taking the first-order approximation yields a linear relationship between the expected phenotypic similarity and the genetic correlation which depends on the heritability. Our approach then entails performing a regression of phenotype similarities between pairs of individuals on their genetic correlation, as estimated from GWAS data. The slope of the regression is then transformed to an estimate of the heritability (Online Methods).


We applied GCR to the same three phenotypes from WTCCC: CD, BD and T1D, and four additional phenotypes: type-2 diabetes (T2D), coronary artery disease (CAD), rheumatoid arthritis (RA) and hypertension (HT). Our method yields considerably higher estimates than the LMM method\citep{Lee2011estimating} for CD and BD (Table \ref{WTCCC_table}).


To explore the nature of this considerable gap between our estimates and previously published estimates of the additive heritability, we conducted a wide range of realistic simulations. We simulated an end-to-end generative model, starting from minor allele frequencies (MAFs) and individual SNP effects, through genotypes and phenotypes, and finally the entire selection process (Online Methods). 

Our simulations differ from all previously reported simulations of case-control studies in the context of heritability estimation in that they yield realistic, rather than degenerate, correlation structure among the individuals in the study. Since we simulated genotypes, the correlation between individuals is not restricted to a small subset of possible values as in the simulations of Lee et al.\citep{Lee2011estimating}. Additionally, when both heritability and ascertainment are high, cases tend to be more genetically similar than expected by chance, which was not accounted for in Lee et al.'s  simulations. In particular, their assumed correlation structure is highly degenerate, with as many as 99.98\% of correlations being exactly 0. For further discussion of the different simulation setups see Online Methods and Supp. Material. 

\begin{figure}
\begin{minipage}[t]{0.45\columnwidth}%
\includegraphics[scale=0.45]{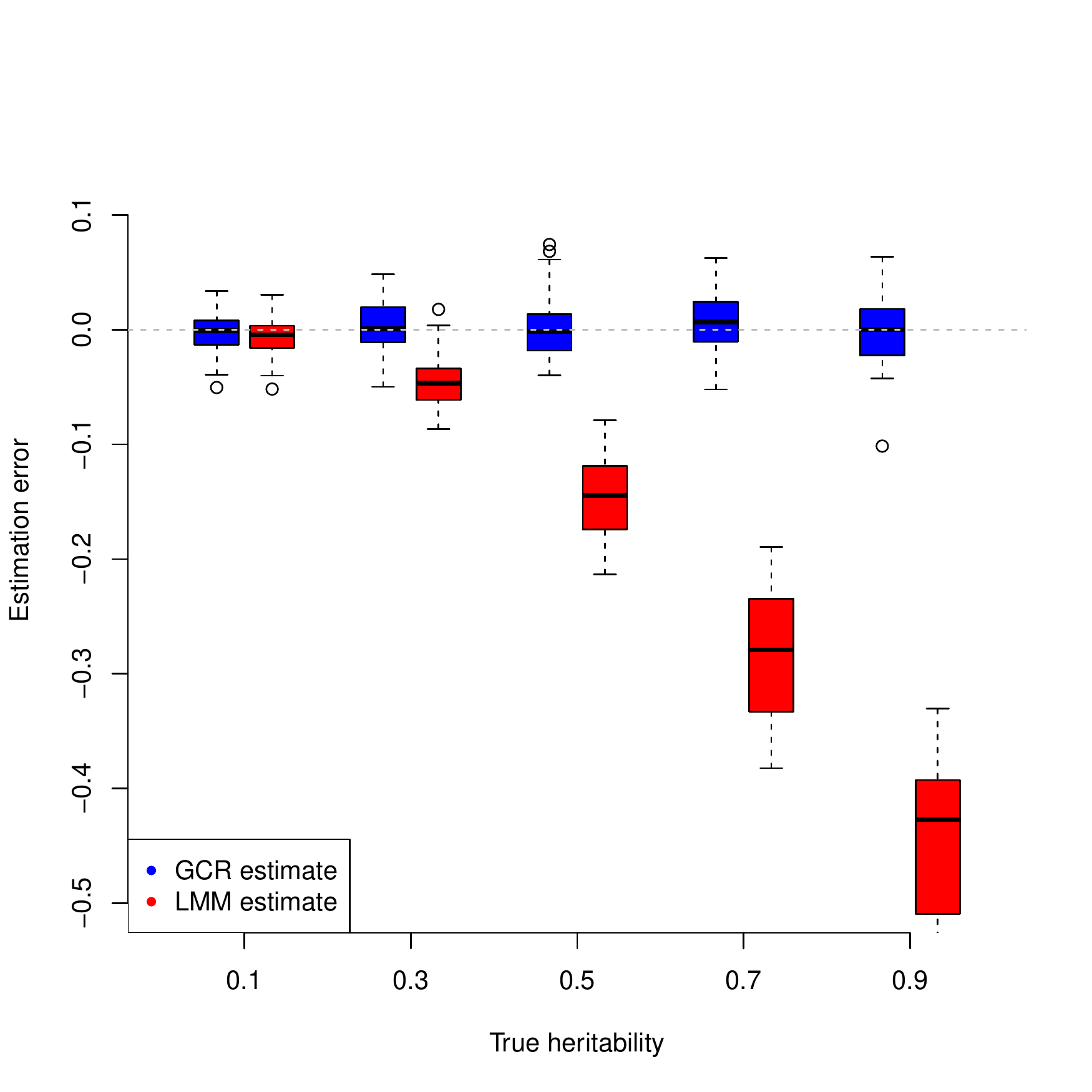}%
\end{minipage}\hfill{}%
\begin{minipage}[t]{0.45\columnwidth}%
\includegraphics[scale=0.45]{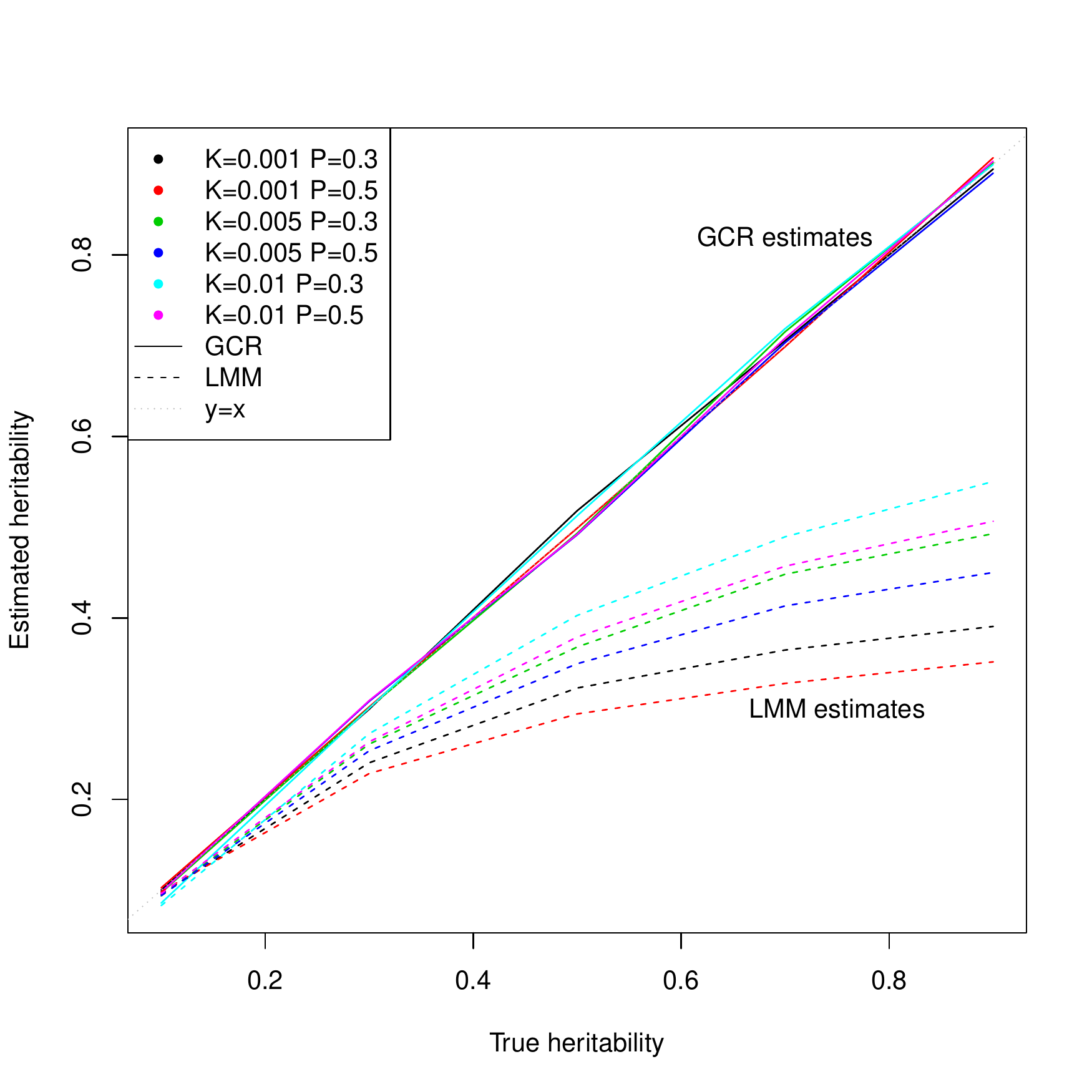}%
\end{minipage}
\caption{Comparing the performance of our GCR approach to the LMM approach of Lee et al.\citep{Lee2011estimating} on realistic simulation setups. K is the disease prevalence, P is the prevalence in the ascertained case-control study, and $h^2$ is the true heritability in the simulation. The left panel shows the distributions of error in estimating heritability as a function of the true heritability for a total of six simulation setups at each heritability level. We observe the significant negative bias incurred by LMM for high heritability values. The right panel shows six example scenarios, and demonstrates the increase in negative bias of LMM as both the true heritability and level of ascertainment increase. At each {(heritability, P, K)} combination we show the average of ten simulation repetitions.}
\label{comparison}
\end{figure}

Our simulations showed that using the LMM method for heritability estimation in case-control studies yielded considerably negatively-biased estimates, while our method consistently generated unbiased estimates (Fig. \ref{comparison} and Supp. Figs. 1-7). 

The simulations also demonstrated a strong relation between the bias of the LMM approach and the increased variance of genetic effects due to ascertainment. This allowed us to derive a heuristic correction for the LMM heritability estimates, which corrected its bias successfully in all our simulation scenarios (Online Methods, Fig. 2 and Supp. Figs. 12-15).
To validate our heuristic correction, we applied it to LMM heritability estimates for all seven WTCCC phenotypes and compared the corrected estimates to the ones obtained using GCR. The correlation between the corrected LMM estimates and GCR estimates was $0.98$, indicating that our heuristic correction generalizes beyond our particular simulations (See Online methods and Supp. material for more details). 

We then applied our correction to the results of several published studies which used the LMM approach. As expected, for studies with low ascertainment or phenotypes with low heritability, the corrected estimates were not substantially different from published estimates (Alzheimer's disease\citep{lee2013estimation}, endometriosis\citep{lee2013estimation}, schizophrenia\citep{lee2012estimating} and Parkinson's disease\citep{do2011web}).
However, the corrected heritability estimate of multiple sclerosis -- the most ascertained study we inspected -- is 52.6\%, compared to the uncorrected estimate of 30\%. See Supp. material for more details. 

An important aspect of heritability estimation is the inclusion of known (fixed) effects that should be accounted for, like known associated SNPs, known environmental effects or sex. However, estimation of fixed effects from ascertained data under the normality assumption might produce biased estimators of both the fixed effects and the heritability\citep{burton2003correcting,bowden2007two,noh2005robust}. Conversely, our method can be rigorously extended to allow for fixed effects (Online Methods). Our simulations suggest that GCR produces accurate estimates of heritability in the presence of fixed effects (see Supp. material for more details).
Specifically for the WTCCC phenotypes, previous analysis\citep{WTCCC}  suggested that there's little population structure (after removing individuals of non British descent). We validated this conclusion using the statistical test described in Patterson et al. (2006)\cite{patterson2006population} (see Supp. material for more details). We therefore included only sex as a fixed effect in our heritability estimation in Table 1. The estimates of both LMM and GCR for most phenotypes without  inclusion of a sex fixed effect are not substantially different from the estimates in Table 1, with the exception of the heart-disease related phenotypes (results without fixed effect not shown). For CAD in particular, it is well known that sex is a major risk factor. Accordingly, the GCR estimate without fixed effect is 71.5\%, dropping to about 61\% once it is added, as seen in Table 1.

GCR is computationally very efficient compared to the LMM approach (running time scales quadratically rather than cubically in the number of individuals), and so the running time is seconds rather than hours on WTCCC cohorts. As the typical size of GWAS continues to increase, this efficiency can prove critical in allowing GCR to remain computationally practical compared to other approaches. Fast running time is also useful in allowing us to use resampling approaches like the jackknife\citep{efron1994introduction} for estimating confidence intervals and standard errors, rather than relying on complex and potentially inaccurate parametric approximations. All standard errors and confidence intervals we report are based on the jackknife.

We also experimented with a more involved version of our approach that included the second order term of the Taylor series expansion. This did not yield more accurate results than our first order GCR in our simulations (Supp. material).

Recently, Zuk et al.\citep{zuk2012mystery} have shown an intriguing general result on the connection between the derivative of the dependence of phenotypic similarity on the proportion of identity by descent (IBD) at the average population IBD, and the narrow-sense heritability, regardless of population structure and phenotype-genotype architecture. 
While Zuk et al.'s result uses IBD and our method relies on estimated genetic correlations, the latter is, in fact, an unbiased estimator of the former, which is unknown. Hence, our method can produce unbiased estimators of narrow-sense heritability, even in the presence of population structure, by multiplying the estimate by a factor of (1-average population IBD).


In conclusion, our new proposed GCR method for estimation of heritability in case-control GWAS, which is based on a regression of phenotype similarities on genotype correlations, is shown to be efficient and accurate. It improves on existing methodology and generates substantially higher estimates of heritability for two major diseases inspected: Crohn's disease and bipolar disorder. Moreover, we provide a heuristic correction for published LMM heritability estimates, which suggests that the heritability of multiple sclerosis is also considerably larger than previously thought. 

\newpage
\noindent{\Large\bf Methods}
~\\
\noindent{\bf Note:} An implementation of our code for GCR estimation, simulations, and heuristic correction of LMM can be accessed at: {\sf https://sites.google.com/site/davidgolanshomepage/gcr}

\section*{Heritability estimation using genetic correlation regression}

{\bf Liability threshold model - notations.} Denote $K$ the prevalence of a condition in the population and $P$
the prevalence in the study. 

Under the liability threshold model, we assume that each individual
$i$ has an unknown liability $l_{i}=g_{i}+e_{i}$ where $g_{i}$
is a genetic random effect, which can be correlated across individuals,
and $e_{i}$ is the environmental random effect, which is assumed
to be independent of each other and of the genetic effects. Both effects
are assumed to follow a Gaussian distribution with variances $\sigma_{g}^{2}$
and $1-\sigma_{g}^{2}$ respectively. A person is then assumed to
be a case if her liability exceeds a threshold $t=\Phi^{-1}(1-K)$,
i.e. the phenotype $y_{i}$ is given by $y_{i}=\mathbb{I}\{l_{i}>t\}$.
This definition guarantees that the prevalence in the population is
indeed $K.$

{\bf Selection probabilities. } When the study is observational, the probability of being included
in the study is independent of the phenotype. However, in a case-control
study, the proportion of cases is usually greatly ascertained. To
model this fact, we define a random indicator variable $s_{i}$ indicating
whether individual $i$ was selected to the study.

The commonly used {}``full'' or {}``complete'' ascertainment assumption\citep{stene1977assumptions} 
is $\mathbb{P}(s_{i}=1\mid y_{i}=1)=1$. While this assumption can be relaxed,
as discussed later, it simplifies subsequent analysis. 

Suppose the population is of size $n$ and that the expected size
of the study is $n_{s}<n$. The expected number of cases in the study
is $nK$. Additionally, the proportion of cases in the study is $P$,
so:
\[
\frac{nK}{n_{s}}=P,\]
yielding:
\[
n_{s}=\frac{nK}{P}.\]
denote $p'$ the probability of a control being included in the study
(i.e. $p'=\mathbb{P}(s_{i}=1\mid y_{i}=0)$). The expected number of controls
in the study is $n(1-K)p'$. Additionally, the proportion of controls
in the study is $(1-P)$ so:
\[
n(1-K)p'=n_{s}(1-P).\]
Solving for $p'$ yields:
\[
p'=\frac{K(1-P)}{P(1-K)}.\]
From here it follows that the probability of being included in the
study for a given individual (with unknown phenotype) is:
\[
K+(1-K)\frac{K(1-P)}{P(1-K)}=\frac{K}{P}.\]
Our results do not depend strictly on the full ascertainment
assumption (that is, $\mathbb{P}(s_{i}=1\mid y_{i}=1)=1$). The latter assumption
can be relaxed such that 
\[
\mathbb{P}(s_{i}=1\mid y_{i}=1)=p^{*},\]
for any $p^{*}$, as long as the probability of being selected as
a control is multiplied by the same probability. This can model any
step prior to the selection procedure, for example the probability
that an individual is approached by the health administration to begin
with. For example, a non-ascertained study ($K=P$) might involve only a proportion
$p^{*}$ of the population. In this case deriving the selection probability
of a given individual yields $p^{*}$ as expected.

{\bf Heritability estimation. } Next, consider a pair of individuals in the study, whose genetic effects
are correlated and denote by $\rho$ the correlation between their genetic effects.

Denote by $Z_{ij}$ the product of the standardized phenotypes:
\[
Z_{ij}=\frac{(y_{i}-P)(y_{j}-P)}{P(1-P)}.\]
The variable $Z_{ij}$ can take three values:
\[
Z_{ij}=\begin{cases}
\frac{1-P}{P}\;\;\;\; & y_{i}=y_{j}=1\\
-1 & y_{i}\ne y_{j}\\
\frac{P}{1-P} & y_{i}=y_{j}=0\end{cases}.\]
We write down the expected value of $Z_{ij}$, conditional on the fact
that $s_{i}=s_{j}=1$ (the individuals are part of the study) and
given $\rho$:
\begin{eqnarray*}
\mathbb{E}[Z_{ij}\mid s_{i}=s_{j}=1;\rho]&=&\frac{1-P}{P}\mathbb{P}(y_{i}=y_{j}=1\mid s_{i}=s_{j}=1;\rho)-\\&&
\mathbb{P}(y_{i}\ne y_{j}\mid s_{i}=s_{j}=1;\rho)+\frac{P}{1-P}\mathbb{P}(y_{i}=y_{j}=0\mid s_{i}=s_{j}=1;\rho).
\end{eqnarray*}
We apply Bayes rule to the first of the three expressions on the right:
\[
\mathbb{P}(y_{i}=y_{j}=1\mid s_{i}=s_{j}=1;\rho)=\frac{\mathbb{P}(s_{i}=s_{j}=1\mid y_{i}=y_{j}=1;\rho)\mathbb{P}(y_{i}=y_{j}=1;\rho)}{\mathbb{P}(s_{i}=s_{j}=1;\rho)}.\]
Under the full ascertainment assumption $\mathbb{P}(s_{i}=s_{j}=1\mid y_{i}=y_{j}=1;\rho)=1$,
and so
\[
\mathbb{P}(y_{i}=y_{j}=1\mid s_{i}=s_{j}=1;\rho)=\frac{\mathbb{P}(y_{i}=y_{j}=1;\rho)}{\mathbb{P}(s_{i}=s_{j}=1;\rho)}.\]
Similarly:
\[
\mathbb{P}(y_{i}=y_{j}=0\mid s_{i}=s_{j}=1;\rho)=\frac{\mathbb{P}(s_{i}=s_{j}=1\mid y_{i}=y_{j}=0;\rho)\mathbb{P}(y_{i}=y_{j}=0;\rho)}{\mathbb{P}(s_{i}=s_{j}=1;\rho)},\]
and since a control is selected to the study with probability $\frac{K(1-P)}{P(1-K)}$,
this boils down to:
\[
\Big(\frac{K(1-P)}{P(1-K)}\Big)^{2}\frac{\mathbb{P}(y_{i}=y_{j}=0;\rho)}{\mathbb{P}(s_{i}=s_{j}=1;\rho)}.\]
For the case of $y_{i}\ne y_{j}$, one individual is a case, and is
automatically selected, while the other is a control and is selected
with probability $\frac{K(1-P)}{P(1-K)}$. Hence:
\[
\mathbb{P}(y_{i}\ne y_{j}\mid s_{i}=s_{j}=1;\rho)=\frac{K(1-P)}{P(1-K)}\frac{\mathbb{P}(y_{i}\ne y_{j};\rho)}{\mathbb{P}(s_{i}=s_{j}=1;\rho)},\]
Using these results we get:
\[
\mathbb{E}[Z_{ij}\mid s_{i}=s_{j}=1;\rho]=\]
\[
\frac{\frac{1-P}{P}\mathbb{P}(y_{i}=y_{j}=1;\rho)-\frac{K(1-P)}{P(1-K)}\mathbb{P}(y_{i}\ne y_{j};\rho)+\frac{P}{1-P}\Big(\frac{K(1-P)}{P(1-K)}\Big)^{2}\mathbb{P}(y_{i}=y_{j}=0;\rho)}{\mathbb{P}(s_{i}=s_{j}=1;\rho)}.\]

Denote the numerator by $A(\rho)$ and the denominator by $B(\rho)$.
We wish to approximate the last equation using a Taylor series around
$\rho=0$. Since the individuals are unrelated, the correlation is expected to be close to 0, and therefore such an approximation is expected to be good.
Such an approximation would take the form:
\[
\mathbb{E}[Z_{ij}\mid s_{i}=s_{j}=1;\rho]\approx\frac{A(0)}{B(0)}+\frac{A'(0)B(0)+B'(0)A(0)}{B(0)^{2}}\rho.\]
See Supp. Mat. for a discussion of a second order approximation.

Note that with $\rho=0$, the phenotypes of the two individuals are
i.i.d. and so $A(0)=0$. Therefore, the Taylor approximation can be
simplified:
\[
\mathbb{E}[Z_{ij}\mid s_{i}=s_{j}=1;\rho]\approx\frac{A'(0)}{B(0)}\rho.\]
Similarly, with $\rho=0$ the events of being included in the study
are i.i.d. for both individuals, so $B(0)=\frac{K^{2}}{P^{2}}$.

All that remains is to find $A'(0)$. Towards that end, we are interested in computing the probabilities of the three possible
combinations of phenotypes:
\begin{eqnarray*}
&&\mathbb{P}(y_{i}=y_{j}=1;\rho,\sigma_{g}^{2})=\int_{t}^{\infty}\int_{t}^{\infty}f_{\rho,\sigma_{g}^{2}}(l_{1},l_{2})dl_{1}dl_{2},\\&&
\mathbb{P}(y_{i}\ne y_{j};\rho,\sigma_{g}^{2})=2\int_{-\infty}^{t}\int_{t}^{\infty}f_{\rho,\sigma_{g}^{2}}(l_{1},l_{2})dl_{1}dl_{2},\\
\mbox{and}&&\\&&
\mathbb{P}(y_{i}=y_{j}=0;\rho,\sigma_{g}^{2})=\int_{-\infty}^{t}\int_{-\infty}^{t}f_{\rho,\sigma_{g}^{2}}(l_{1},l_{2})dl_{1}dl_{2},\end{eqnarray*}
where $f_{\rho,\sigma_{g}^{2}}$ is the multivariate Gaussian density,
namely:
\[
f_{\rho,\sigma_{g}^{2}}(l_{1},l_{2})=\frac{1}{2\pi}|\Sigma|^{-\frac{1}{2}}e^{-\frac{(l_{1},l_{2})\Sigma^{-1}(l_{1},l_{2})^{\intercal}}{2}},\]
with $\Sigma$ denoting the covariance matrix of the liabilities,
given explicitly by: \[
\Sigma=\left(\begin{array}{cc}
1 & \rho\\
\rho & 1\end{array}\right)\sigma_{g}^{2}+\left(\begin{array}{cc}
1 & 0\\
0 & 1\end{array}\right)(1-\sigma_{g}^{2})=\left(\begin{array}{cc}
1 & \rho\sigma_{g}^{2}\\
\rho\sigma_{g}^{2} & 1\end{array}\right).\]
The determinant of $\Sigma$ is $|\Sigma|=1-\rho^{2}\sigma_{g}^{4}$
and its inverse is $\Sigma^{-1}=\frac{1}{1-\rho^{2}\sigma_{g}^{4}}\left(\begin{array}{cc}
1 & -\rho\sigma_{g}^{2}\\
-\rho\sigma_{g}^{2} & 1\end{array}\right),$ and so the density function $f_{\rho,\sigma_{g}^{2}}$ can be written
as:
\[
f_{\rho,\sigma_{g}^{2}}(l_{1},l_{2})=\frac{1}{2\pi\sqrt{1-\rho^{2}\sigma_{g}^{4}}}e^{-\frac{l_{1}^{2}+l_{2}^{2}-2l_{1}l_{2}\rho\sigma_{g}^{2}}{2(1-\rho^{2}\sigma_{g}^{4})}}.\]
Differentiating $A(\rho)$ requires differentiating each of the three double integrals
w.r.t. $\rho$:
\[
\frac{d}{d\rho}\int_{t}^{\infty}\int_{t}^{\infty}f_{\rho,\sigma_{g}^{2}}(l_{1},l_{2})dl_{1}dl_{2}=\int_{t}^{\infty}\int_{t}^{\infty}\frac{d}{d\rho}f_{\rho,\sigma_{g}^{2}}(l_{1},l_{2})dl_{1}dl_{2}.\]
Setting $\rho=0$ in the last expression yields:
\[
\int_{t}^{\infty}\int_{t}^{\infty}l_{1}l_{2}\sigma_{g}^{2}\frac{1}{2\pi}e^{-\frac{l_{1}^{2}+l_{2}^{2}}{2}}=\sigma_{g}^{2}\Big[\int_{t}^{\infty}l\frac{1}{\sqrt{2\pi}}e^{-\frac{l^{2}}{2}}dl\Big]^{2}=\sigma_{g}^{2}\varphi(t)^{2}.\]
Explanation: we differentiate and set $\rho=0$. By the chain rule, the derivative
of any expression with $\rho^{2}$ is $0$ at $\rho=0$, and obviously
the derivative of any expression which does not depend on $\rho$
is $0$. The only expression whose derivative is therefore not $0$
at $\rho=0$ is $-2l_{1}l_{2}\rho\sigma_{g}^{2}$ in the numerator
of the exponent. The denominator of the exponent is $2$ at $\rho=0$,
and so the derivative at $\rho=0$ is $l_{1}l_{2}\sigma_{g}^{2}\frac{1}{2\pi}e^{-\frac{l_{1}^{2}+l_{2}^{2}}{2}}$.
Similarly:
\[
\frac{d}{d\rho}\int_{-\infty}^{t}\int_{t}^{\infty}f_{\rho,\sigma_{g}^{2}}(l_{1},l_{2})dl_{1}dl_{2}=-\sigma_{g}^{2}\varphi(t)^{2},\]
and 
\[
\frac{d}{d\rho}\int_{-\infty}^{t}\int_{-\infty}^{t}\, f_{\rho,\sigma_{g}^{2}}(l_{1},l_{2})dl_{1}dl_{2}=\sigma_{g}^{2}\varphi(t)^{2}.\]
Using these results we can write down $A'(0)$:
\[
A'(0)=\Big[\frac{1-P}{P}+2\frac{K(1-P)}{P(1-K)}+\frac{P}{1-P}\Big(\frac{K(1-P)}{P(1-K)}\Big)^{2}\Big]\sigma_{g}^{2}\varphi(t)^{2}=\frac{1-P}{P(1-K)^{2}}\sigma_{g}^{2}\varphi(t)^{2},\]
and so:
\[
\mathbb{E}[Z_{ij}\mid s_{i}=s_{j}=1;\rho]\approx\frac{A'(0)}{B(0)}\rho=\frac{\frac{1-P}{P(1-K)^{2}}\sigma_{g}^{2}\varphi(t)^{2}}{\frac{K^{2}}{P^{2}}}\rho=\frac{P(1-P)}{K^{2}(1-K)^{2}}\sigma_{g}^{2}\varphi(t)^{2}\rho.\]
Hence, when the error of the approximation is small, the slope obtained
by regressing $Z_{ij}$ on $G_{ij}$ is an unbiased estimator of $\frac{P(1-P)}{K^{2}(1-K)^{2}}\sigma_{g}^{2}\varphi(t)^{2}$,
thus dividing it by $\frac{P(1-P)}{K^{2}(1-K)^{2}}\varphi(t)^{2}$
yields an unbiased estimator of $\sigma_{g}^{2}$ - the liability
scale heritability. 

{\bf Extending the liability threshold model to include fixed effects.} It is often desired to include fixed effects in the analysis of a
complex phenotype. Such fixed effects might include external information
such as sex, diet and exposure to environmental risks, but can also
be genetic variants with known effects or estimates of population
structure such as projections of several top principal components.

Since the liability threshold model is in fact a probit model, fixed effects
can be included in the usual manner:
\[l_{i}=x_{i}^{\intercal}\beta+g_{i}+e_{i},\]
where 
$x_{i}$ is a vector of the values of the relevant covariates
and $\beta$ is a vector of their respective effect sizes. 

An individual is a case if $l_{i}>t$, as before. However, an equivalent
formulation would be to subtract the fixed effect from the threshold,
rather than adding it to the liability:
\[t_{i}=t-x_{i}^{\intercal}\beta,\]
thus keeping the previous formulation of the liability as a sum of
genetic and environmental effects.

{\bf Heritability estimation with known fixed effects.} Assume first that the fixed effects are known, and so the $t_{i}$'s are known. The probability of being a case, and of being included in the study, are no longer equal for all the observed individuals. We denote:
\[
K_{i}=\mathbb{P}(y_{i}=1;t_{i}),\]
the probability that the $i$'th individual is a case, and:
\[
P_{i}=\mathbb{P}(y_{i}=1\mid s_{i}=1;t_{i}),\]
the probability that the $i$'th individual is a case conditional on being selected for the study, where $t_i$ is computed using the fixed effects as described above. 
We now wish to derive the same first order approximation, while accounting for the newly introduced heterogeneity. We redefine:
\[
Z_{ij}=\frac{(y_{i}-P_{i})(y_{j}-P_{j})}{\sqrt{P_{i}(1-P_{i})}\sqrt{P_{j}(1-P_{j})}},\]
and follow the same steps as before (see supplementary materials for a full derivation). Using the first order Taylor approximation we conclude that the slope obtained by regressing $Z_{ij}$ on 
\[
\frac{\varphi(t_{i})\varphi(t_{j})\Big[1-(P_{i}+P_{j})\Big(\frac{P-K}{P(1-K)}\Big)+P_{i}P_{j}\Big(\frac{P-K}{P(1-K)}\Big)^{2}\Big]}{\sqrt{P_{i}(1-P_{i})}\sqrt{P_{j}(1-P_{j})}\Big(K_{i}+(1-K_{i})\frac{K(1-P)}{P(1-K)}\Big)\Big(K_{j}+(1-K_{j})\frac{K(1-P)}{P(1-K)}\Big)}G_{ij},\]
is an estimator of heritability on the liability scale.

{\bf Estimating heritability with unknown fixed effects.} More often than not, the effects of relevant fixed effects are unknown
and must be estimated from the data. However, estimating effect sizes
under ascertainment in case-control studies is notoriously problematic.
Specifically, under the threshold (probit) model, ignoring the ascertainment
yields biased estimators.

A special exception is the case of logistic regression. In their seminal
paper, Prentice and Pyke (1979)\citep{prentice1979logistic} proved that using a logistic regression to
estimate fixed effects from ascertained data yields consistent estimators
of these effects in the (unascertained) population, and that the ascertainment
only biases the intercept. 

We therefore suggest a two-step procedure for estimating heritability.
First, we estimate the fixed effects using a logistic regression model.
We then correct the effect of the ascertainment, and obtain the individual-specific
thresholds. Lastly, we plug the thresholds into the estimation scheme
described above. 

More elaborately, by Bayes' formula:
\[P_{i}=\frac{\mathbb{P}(s_{i}=1\mid y_{i}=1;x_{i})K_i}{\mathbb{P}(s_{i}=1 ; x_{i})},\]
by the complete ascertainment assumption $\mathbb{P}(s_{i}=1\mid y_{i}=1,x_{i})=1$,
and according to the selection scheme:
\[\mathbb{P}(s_{i}=1 ; x_{i})=K_{i}+\frac{K(1-P)}{P(1-K)}\big(1-K_{i}\big).\]
We can thus solve for $K_i$ and express it is a function of $P_i$:
\[
K_{i} =  \frac{\frac{K(1-P)}{P(1-K)} P_{i}}{1 + \frac{K(1-P)}{P(1-K)} P_{i} - P_{i}}  
\]        

We then use logistic regression to obtain  $\hat{P}_{i}$ - a consistent estimator of $P_{i}$, and use this estimate to obtain an estimate of $K_i$, which is in turn used to estimate the threshold: 
\[\hat{t}_{i}=\Phi^{-1}\big(1-\hat{K}_{i}\big),\]
and the estimates of the individual-wise thresholds are used for estimating
the variance of the genetic effect.

{\bf Estimating the added variance due to fixed effects}
Lastly, the presence of fixed effects increases the variance of the liability, so $\sigma_{g}^{2}$ no longer equals $h^2$. The appropriate definition of heritability is now: 
\[
h^2=\frac{\sigma_{g}^{2}}{\sigma_{g}^{2}+\sigma_{e}^{2}+\sigma_{t}^{2}}=\frac{\sigma_{g}^{2}}{1+\sigma_{t}^{2}},
\]
where $\sigma_{t}^{2}$ is the variance of the thresholds in the population, and so  the estimate of $\sigma_{g}^{2}$ can be transformed to an estimate of the heritability simply by dividing it by  $1+\sigma_{t}^{2}$. We discuss how $\sigma_{t}^{2}$ can be estimated from the data in the Supplementary material.  

\section*{Simulations using a generative model}

Our goal was to create a realistic simulation, covering a wide range of combinations of disease prevalence $K$, case sampling probability $P$ and heritability $\sigma^2_g$, and for each one creating ``natural'' genotypes that realistically recreate the complex correlation structure induced by case-control sampling. 

Given these parameters, our simulations proceeded as follows:
\begin{enumerate}
\item The MAFs of 10,000 SNPs were randomly sampled from $U[0.05,0.5]$. 
\item SNP effect sizes were randomly sampled from $N(0,\frac{\sigma_{g}^{2}}{m}).$
\item For each individual, we:
\begin{enumerate}
\item Randomly generated a genotype using the MAFs, and normalized it (according
to Yang et al.'s model\citep{yang2011gcta}).
\item Used the genotype and the effect sizes, to compute the genetic effect.
\item Sampled an environmental effect from $N(0,1-\sigma_{g}^{2})$.
\item Computed liability and phenotype.
\item If the phenotype was a case - the individual was automatically included
in the study. Otherwise the individual was included in the study with
probability $\frac{K(1-P)}{P(1-K)}$. 
\end{enumerate}
\item Step (2) was repeated until enough individuals were accumulated (4,000). 
\item The genotypes of all included individuals were used to compute $G=\frac{ZZ^{\intercal}}{m}$
where $Z$ is the matrix of n5.8alized genotypes. 
\end{enumerate}
This matrix $G$ was then used to estimate heritability for the LMM using the $GCTA$ software\citep{yang2011gcta} and the correction of Lee et al.\citep{Lee2011estimating}; and using GCR as described above.

We note that our choice to work with SNPs that are in linkage equilibrium was motivated by the analysis of Patterson et al.\cite{patterson2006population}. They show that for the purpose of generating correlation matrices, using SNPs in linkage disequilibrium (LD) is equivalent to using a smaller number of SNPs in linkage equilibrium. They also suggest a method for estimating the effective number of SNPs (i.e. the number of SNPs in linkage equilibrium leading to the same distribution of correlation matrices as a given set of SNPs in LD). Applying their method to WTCCC data suggests that the effective number of SNPs in linkage equilibrium is roughly one tenth of the actual number of SNPs. Hence, using 10,000 SNPs in our simulations is equivalent to simulating roughly 100,000 SNPs with realistic LD structure.

\section*{Heuristic correction for the LMM approach}

Denote the estimate of the heritability on the liability scale, obtained through the method of Lee et al.\citep{Lee2011estimating} by $\tilde{\sigma}^2_g$, and the estimates obtained by our method by  $\hat{\sigma}^2_g$. Denote the true variance of the genetic effect under ascertainment by: 
$$\sigma_{g_{cc}}^{2}=\sigma_{g}^{2}\Big[1+\sigma_{g}^{2}\varphi(t)\frac{(P-K)}{K(1-K)}\Big[t-\varphi(t)\frac{(P-K)}{K(1-K)}\Big]\Big].$$
We note a different analytical expression of $\sigma_{g_{cc}}^{2}$
is given in\citep{Lee2011estimating}. We validated correctness of our derivation of this expression numerically.

As detailed in the Supplementary Materials, our simulations demonstrated the following properties of these estimators:
\begin{enumerate}
\item They are both unbiased when there is no ascertainment.
\item Our estimate $\hat{\sigma}^2_g$ remains unbiased in all situations. However, in presence of ascertainment,  $\tilde{\sigma}^2_g$ is biased, and is not linear in the true $\sigma^2_g$ for fixed $K,P$. 
\item When multiplied by $\sigma_{g_{cc}}^{2}/\sigma_{g}^{2}$, the estimate $\tilde{\sigma}^2_g$ becomes linear in $\sigma^2_g$ for fixed $K,P$.  
\item The bias of $\tilde{\sigma}^2_g$ worsens as the ascertainment factor $\frac{K}{P}$ decreases.
\end{enumerate}
We therefore performed an extensive analysis of the relationship between $\frac{K}{P}$ and the bias of the ``linearized'' estimate $\tilde{\sigma}^2_g \frac{\sigma_{g_{cc}}^{2}}{\sigma_{g}^{2}}$ in our simulations. Our analysis, as detailed in Supplementary Materials, led us to the following relationship between $\tilde{\sigma}^2_g$ and the true underlying heritability $\sigma_{g}^{2}$:

$$\mathbb{E}\tilde{\sigma}^{*2}_g \approx \frac{K^2(1-K)^2}{\varphi(t)^2 P (1-P)} (1.3-0.3\sqrt{\frac{K}{P}}) \frac{\sigma_{g}^{4}}{\sigma_{g_{cc}}^{2}}.
$$

{\bf Correcting published results}
We have derived our heuristic correction using simulations
wherein the true underlying heritability is known. However, contrary to previously used corrections, our correction is not a linear transformation of the estimate.
When attempting to correct published estimates, we only know $K,P$ and $\tilde{\sigma}_{g}^{2}$.
We define our corrected estimate $\hat{\sigma}_{g}^{*2}$ to be
the value of $\sigma_{g}^{2}$ for which \[
\mathbb{E}[\tilde{\sigma}_{g}^{2};\sigma_{g}^{2}=\hat{\sigma}_{g}^{*2}]=\tilde{\sigma}_{g}^{2},\]
where the expectation is computed using the approximate relationship we derived previously. 
In other words, the corrected estimate of heritability is the value of heritability for which the expectation of the estimator is the observed estimate, where the expectation is
calculated using our heuristic correction. 

Confidence intervals are derived by applying the same procedure to
the top and bottom limits of a $95\%$ confidence interval based on
the published standard deviation of the estimate.

\vspace{2in}


\bibliographystyle{nature}
\bibliography{main}

\begin{table}
\begin{tabular}{|c|c||c|c||c|}
\hline 
Phenotype & Prevalence (\%) & LMM (sd) (\%)& GCR (sd) (\%)& Family studies (\%)\tabularnewline
\hline
\hline 
CD & 0.1 & 23.2 (3) & 34.1 (5.8) & 50-60 \tabularnewline
\hline
 BD & 0.5 & 42.8 (4.1) & 53.8 (6.8) & 71 \tabularnewline
\hline
 T2D & 3 & 42 (6.3) & 47.8 (9.9) & 26-69 \tabularnewline
\hline
 HT & 5 & 53.1 (7.4) & 52.3 (10.6) & 31-63** \tabularnewline
\hline
 CAD & 30 & 66.9 (12.8) & 61.1 (16.9) & 39-56** \tabularnewline
\hline
 RA & 0.75 & 16.5 (4.6) & 17.9 (7) & 53-65 \tabularnewline
\hline
 T1D* & 0.5 & 16.1 (4.2) & 17.2 (5.8) & 72-88 \tabularnewline
\hline

\end{tabular}
\caption{Comparing estimated heritability of 7 WTCCC studies using GCR, LMM and family-based estimates. Standard deviations are given in parentheses. GCR and LMM estimates are adjusted for imperfect LD as discussed in \citep{yang2010common}, and include sex as a fixed effect. GCR standard deviations are estimated using 1,000 jackknife iterations and LMM standard deviations are produced by the GCTA software\citep{yang2011gcta}. Population prevalence and family-based estimates are from \citep{Lee2011estimating,wray2010genetic,dudbridge2013power,almgren2011heritability,poulsen1999heritability,sofaer1993crohn,edvardsen2008heritability,kyvik1995concordance,hyttinen2003genetic,23andme,WTCCC,bhf,kupper2005heritability}.
\newline
$*$ Analysis does not include chromosome 6.
\newline
$**$ Estimates might not be comparable to GCR/LMM estimates due to different phenotype definitions.
}

\label{WTCCC_table}
\end{table}

\begin{figure}
\begin{minipage}[t]{0.45\columnwidth}%
\includegraphics[scale=0.45]{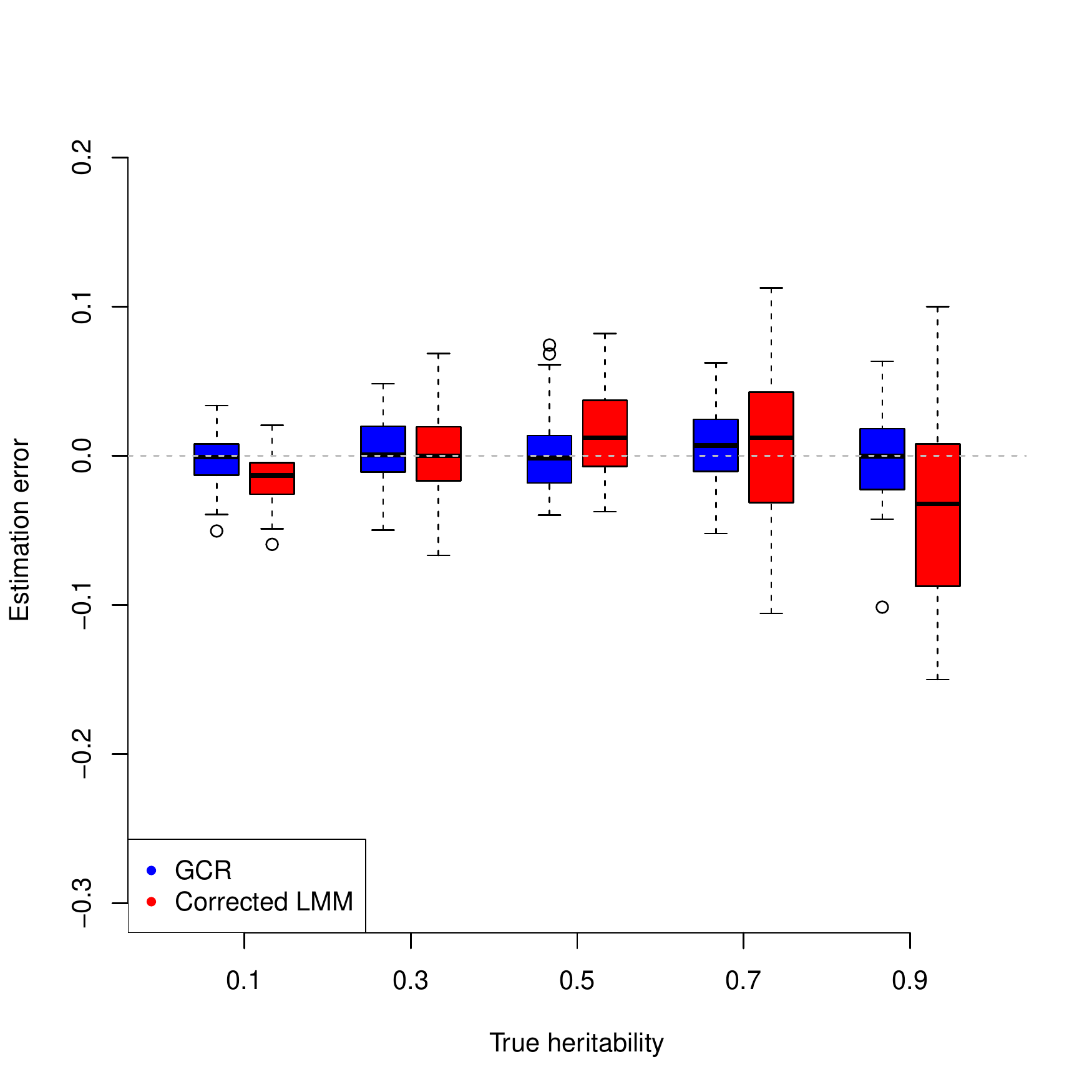}%
\end{minipage}\hfill{}%
\begin{minipage}[t]{0.45\columnwidth}%
\includegraphics[scale=0.45]{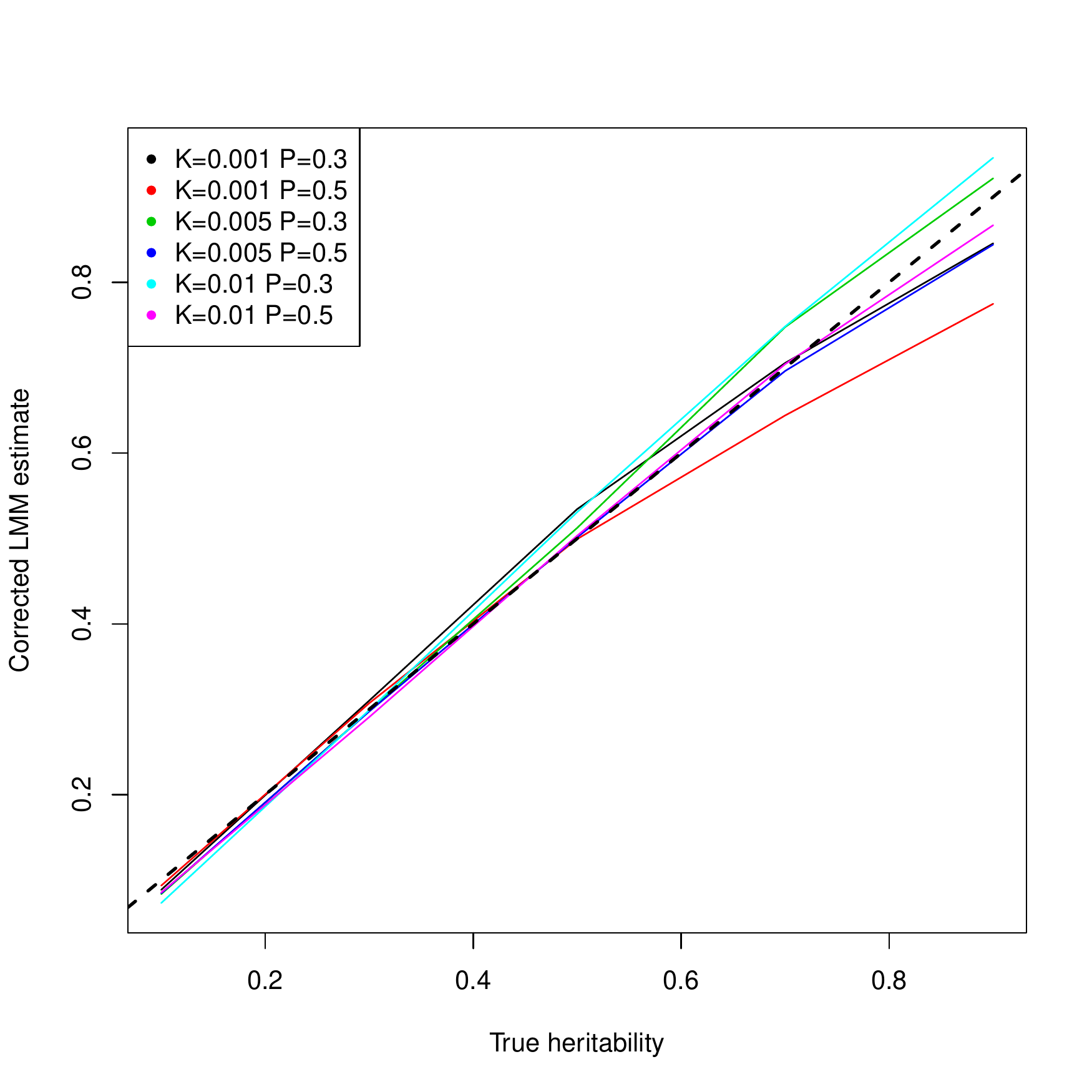}%
\end{minipage}
\caption{Heuristic correction of LMM estimates. Left - comparison of estimation errors of GCR and corrected LMM estimates, using the same simulations as in Figure \ref{comparison}.  Right - breakdown of the average  corrected estimates by K,P and $h^2$. We see that applying the heuristic correction yields estimates which are still much noisier than GCR estimates, but are considerably less biased than LMM estimates. The heuristic correction works best for intermediate values of the underlying heritability.}

\label{correction_figure}

\end{figure}

\end{document}


\title{Narrowing the gap on heritability of common disease by direct estimation in case-control GWAS \\ {\it Supplementary Materials} }
\date{April 2013}

\author{David Golan$^1$ \& Saharon Rosset$^1$}

\maketitle

\begin{enumerate}
 \item Department of Statistics and Operations Research, Tel-Aviv University.
 \end{enumerate}

\noindent{\bf Note:} This supplementary text includes a complete and self-contained description of our methods, repeating the material from Online Methods where appropriate. 

\tableofcontents{}
\pagebreak

\section{Heritability estimation using genetic correlation regression}

\subsection{Liability threshold model - notations}

Denote $K$ the prevalence of a condition in the population and $P$
the prevalence in the study. 

Under the liability threshold model, we assume that each individual
$i$ has an unknown liability $l_{i}=g_{i}+e_{i}$ where $g_{i}$
is a genetic random effect, which can be correlated across individuals,
and $e_{i}$ is the environmental random effect, which is assumed
to be independent of each other and of the genetic effects. Both effects
are assumed to follow a Gaussian distribution with variances $\sigma_{g}^{2}$
and $1-\sigma_{g}^{2}$ respectively. A person is then assumed to
be a case if her liability exceeds a threshold $t=\Phi^{-1}(1-K)$,
i.e. the phenotype $y_{i}$ is given by $y_{i}=\mathbb{I}\{l_{i}>t\}$.
This definition guarantees that the prevalence in the population is
indeed $K.$

\subsection{Selection probabilities }

When the study is observational, the probability of being included
in the study is independent of the phenotype. However, in a case-control
study, the proportion of cases is usually greatly ascertained. To
model this fact, we define a random indicator variable $s_{i}$ indicating
whether individual $i$ was selected to the study.

The commonly used {}``full ascertainment'' \citep{stene1977assumptions} assumption
is $P(s_{i}=1\mid y_{i}=1)=1$. While this assumption can be relaxed,
as discussed later, it simplifies subsequent analysis. 

Suppose the population is of size $n$ and that the expected size
of the study is $n_{s}<n$. The expected number of cases in the study
is $nK$. Additionally, the proportion of cases in the study is $P$,
so:

\[
\frac{nK}{n_{s}}=P,\]

yielding:

\[
n_{s}=\frac{nK}{P}.\]

denote $p'$ the probability of a control being included in the study
(i.e. $p'=P(s_{i}=1\mid y_{i}=0)$). The expected number of controls
in the study is $n(1-K)p'$. Additionally, the proportion of controls
in the study is $(1-P)$ so:

\[
n(1-K)p'=n_{s}(1-P).\]

Solving for $p'$ yields:

\[
p'=\frac{K(1-P)}{P(1-K)}.\]

From here it follows that the probability of being included in the
study for a given individual (with unknown phenotype) is:

\[
K+(1-K)\frac{K(1-P)}{P(1-K)}=\frac{K}{P}.\]

\subsubsection{Relaxing the {}``full ascertainment'' assumption}

In fact, the latter results do not depend strictly on the full ascertainment
assumption (that is, $P(s_{i}=1\mid y_{i}=1)=1$). The latter assumption
can be relaxed such that 

\[
P(s_{i}=1\mid y_{i}=1)=p^{*},\]

for any $p^{*}$, as long as the probability of being selected as
a control is multiplied by the same probability. This can model any
step prior to the selection procedure, for example the probability
that an individual is approached by the health administration to begin
with. For example, a non-ascertained study might involve only a proportion
$p^{*}$of the population. In this case deriving the selection probability
of a given individual yields $p^{*}$ as expected.

\subsection{Heritability estimation }

Next, consider a pair of individuals in the study, whose genetic effects
are correlated and denote by $\rho$ the correlation.

Denote by $Z_{ij}$ the product of the standardized phenotypes:

\[
Z_{ij}=\frac{(y_{i}-P)(y_{j}-P)}{P(1-P)}.\]

The variable $Z_{ij}$ can obtain three values:

\[
Z_{ij}=\begin{cases}
\frac{1-P}{P}\;\;\;\; & y_{i}=y_{j}=1\\
-1 & y_{i}\ne y_{j}\\
\frac{P}{1-P} & y_{i}=y_{j}=0\end{cases}.\]

We write down the expected value f $Z_{ij}$, conditional on the fact
that $s_{i}=s_{j}=1$ (the individuals are part of the study) and
given $\rho$:

\[
\mathbb{E}[Z_{ij}\mid s_{i}=s_{j}=1;\rho]=\frac{1-P}{P}\mathbb{P}(y_{i}=y_{j}=1\mid s_{i}=s_{j}=1;\rho)-\]

\[
\mathbb{P}(y_{i}\ne y_{j}\mid s_{i}=s_{j}=1;\rho)+\frac{P}{1-P}\mathbb{P}(y_{i}=y_{j}=0\mid s_{i}=s_{j}=1;\rho).\]

We apply Bayes rule to the first of the three summands on the right:

\[
\mathbb{P}(y_{i}=y_{j}=1\mid s_{i}=s_{j}=1;\rho)=\frac{\mathbb{P}(s_{i}=s_{j}=1\mid y_{i}=y_{j}=1;\rho)\mathbb{P}(y_{i}=y_{j}=1;\rho)}{\mathbb{P}(s_{i}=s_{j}=1;\rho)}.\]

Under the full ascertainment assumption $\mathbb{P}(s_{i}=s_{j}=1\mid y_{i}=y_{j}=1;\rho)=1$,
and so

\[
\mathbb{P}(y_{i}=y_{j}=1\mid s_{i}=s_{j}=1;\rho)=\frac{\mathbb{P}(y_{i}=y_{j}=1;\rho)}{\mathbb{P}(s_{i}=s_{j}=1;\rho)}.\]

Similarly:

\[
\mathbb{P}(y_{i}=y_{j}=0\mid s_{i}=s_{j}=1;\rho)=\frac{\mathbb{P}(s_{i}=s_{j}=1\mid y_{i}=y_{j}=0;\rho)\mathbb{P}(y_{i}=y_{j}=0;\rho)}{\mathbb{P}(s_{i}=s_{j}=1;\rho)},\]

and since a control is selected to the study with probability $\frac{K(1-P)}{P(1-K)}$,
this boils down to:

\[
\Big(\frac{K(1-P)}{P(1-K)}\Big)^{2}\frac{\mathbb{P}(y_{i}=y_{j}=0;\rho)}{\mathbb{P}(s_{i}=s_{j}=1;\rho)}.\]

For the case of $y_{i}\ne y_{j}$, one individual is a case, and is
automatically selected, while the other is a control and is selected
with probability $\frac{K(1-P)}{P(1-K)}$. Hence:

\[
\mathbb{P}(y_{i}\ne y_{j}\mid s_{i}=s_{j}=1;\rho)=\frac{K(1-P)}{P(1-K)}\frac{\mathbb{P}(y_{i}\ne y_{j};\rho)}{\mathbb{P}(s_{i}=s_{j}=1;\rho)},\]

Using these results we get:

\[
\mathbb{E}[Z_{ij}\mid s_{i}=s_{j}=1;\rho]=\]

\[
\frac{\frac{1-P}{P}\mathbb{P}(y_{i}=y_{j}=1;\rho)-\frac{K(1-P)}{P(1-K)}\mathbb{P}(y_{i}\ne y_{j};\rho)+\frac{P}{1-P}\Big(\frac{K(1-P)}{P(1-K)}\Big)^{2}\mathbb{P}(y_{i}=y_{j}=0;\rho)}{\mathbb{P}(s_{i}=s_{j}=1;\rho)}.\]

Denote the numerator by $A(\rho)$ and the denominator by $B(\rho)$.
We wish to approximate the latter equation using a Taylor series around
$\rho=0$. Such an approximation would take the form:

\[
\mathbb{E}[Z_{ij}\mid s_{i}=s_{j}=1;\rho]\approx\frac{A(0)}{B(0)}+\frac{A'(0)B(0)+B'(0)A(0)}{B(0)^{2}}\rho.\]
We later discuss a second order approximation as well. 

Note that with $\rho=0$, the phenotypes of the two individuals are
i.i.d. and so $A(0)=0$. Therefore, the Taylor approximation can be
simplified:

\[
\mathbb{E}[Z_{ij}\mid s_{i}=s_{j}=1;\rho]\approx\frac{A'(0)}{B(0)}\rho.\]

Similarly, with $\rho=0$ the events of being included in the study
are i.i.d. for both individuals, so $B(0)=\frac{K^{2}}{P^{2}}$.

All that remains is to find $A'(0)$.

We are interested in computing the probabilities of the three possible
combinations of phenotypes:

\[
P(y_{i}=y_{j}=1;\rho,\sigma_{g}^{2})=\int_{t}^{\infty}\int_{t}^{\infty}f_{\rho,\sigma_{g}^{2}}(l_{1},l_{2})dl_{1}dl_{2},\]

\[
P(y_{i}\ne y_{j};\rho,\sigma_{g}^{2})=2\int_{-\infty}^{t}\int_{t}^{\infty}f_{\rho,\sigma_{g}^{2}}(l_{1},l_{2})dl_{1}dl_{2},\]

and

\[
P(y_{i}=y_{j}=0;\rho,\sigma_{g}^{2})=\int_{-\infty}^{t}\int_{-\infty}^{t}f_{\rho,\sigma_{g}^{2}}(l_{1},l_{2})dl_{1}dl_{2},\]

where $f_{\rho,\sigma_{g}^{2}}$ is the multivariate Gaussian density,
namely:

\[
f_{\rho,\sigma_{g}^{2}}(l_{1},l_{2})=\frac{1}{2\pi}|\Sigma|^{-\frac{1}{2}}e^{-\frac{(l_{1},l_{2})\Sigma^{-1}(l_{1},l_{2})^{\intercal}}{2}},\]

with $\Sigma$ denoting the covariance matrix of the liabilities,
given explicitly by: \[
\Sigma=\left(\begin{array}{cc}
1 & \rho\\
\rho & 1\end{array}\right)\sigma_{g}^{2}+\left(\begin{array}{cc}
1 & 0\\
0 & 1\end{array}\right)(1-\sigma_{g}^{2})=\left(\begin{array}{cc}
1 & \rho\sigma_{g}^{2}\\
\rho\sigma_{g}^{2} & 1\end{array}\right).\]

The determinant of $\Sigma$ is $|\Sigma|=1-\rho^{2}\sigma_{g}^{4}$
and its inverse is $\Sigma^{-1}=\frac{1}{1-\rho^{2}\sigma_{g}^{4}}\left(\begin{array}{cc}
1 & -\rho\sigma_{g}^{2}\\
-\rho\sigma_{g}^{2} & 1\end{array}\right),$ and so the density function $f_{\rho,\sigma_{g}^{2}}$ can be written
as:

\[
f_{\rho,\sigma_{g}^{2}}(l_{1},l_{2})=\frac{1}{2\pi\sqrt{1-\rho^{2}\sigma_{g}^{4}}}e^{-\frac{l_{1}^{2}+l_{2}^{2}-2l_{1}l_{2}\rho\sigma_{g}^{2}}{2(1-\rho^{2}\sigma_{g}^{4})}}.\]

Deriving $A(\rho)$ requires deriving each of the three double integrals
w.r.t. $\rho$:

\[
\frac{d}{d\rho}\int_{t}^{\infty}\int_{t}^{\infty}f_{\rho,\sigma_{g}^{2}}(l_{1},l_{2})dl_{1}dl_{2}=\int_{t}^{\infty}\int_{t}^{\infty}\frac{d}{d\rho}f_{\rho,\sigma_{g}^{2}}(l_{1},l_{2})dl_{1}dl_{2}.\]

Setting $\rho=0$ in the last expression yields:

\[
\int_{t}^{\infty}\int_{t}^{\infty}l_{1}l_{2}\sigma_{g}^{2}\frac{1}{2\pi}e^{-\frac{l_{1}^{2}+l_{2}^{2}}{2}}=\sigma_{g}^{2}\Big[\int_{t}^{\infty}l\frac{1}{\sqrt{2\pi}}e^{-\frac{l^{2}}{2}}dl\Big]^{2}=\sigma_{g}^{2}\varphi(t)^{2}.\]

Explanation: we differentiate and set $\rho=0$. By the chain rule, the derivative
of any expression with $\rho^{2}$ is $0$ with $\rho=0$, and obviously
the derivative of any expression which does not depend on $\rho$
is $0$. The only expression whose derivative is therefore not $0$
at $\rho=0$ is $-2l_{1}l_{2}\rho\sigma_{g}^{2}$ in the numerator
of the exponent. The denominator of the exponent is $2$ at $\rho=0$,
and so the derivative is $l_{1}l_{2}\sigma_{g}^{2}\frac{1}{2\pi}e^{-\frac{l_{1}^{2}+l_{2}^{2}}{2}}$.

Similarly:

\[
\frac{d}{d\rho}\int_{-\infty}^{t}\int_{t}^{\infty}f_{\rho,\sigma_{g}^{2}}(l_{1},l_{2})dl_{1}dl_{2}=-\sigma_{g}^{2}\varphi(t)^{2},\]

and 

\[
\frac{d}{d\rho}\int_{-\infty}^{t}\int_{-\infty}^{t}\, f_{\rho,\sigma_{g}^{2}}(l_{1},l_{2})dl_{1}dl_{2}=\sigma_{g}^{2}\varphi(t)^{2}.\]

Using these results we can write down $A'(0)$:

\[
A'(0)=\Big[\frac{1-P}{P}+2\frac{K(1-P)}{P(1-K)}+\frac{P}{1-P}\Big(\frac{K(1-P)}{P(1-K)}\Big)^{2}\Big]\sigma_{g}^{2}\varphi(t)^{2}=\frac{1-P}{P(1-K)^{2}}\sigma_{g}^{2}\varphi(t)^{2},\]

and so:

\[
\mathbb{E}[Z_{ij}\mid s_{i}=s_{j}=1;\rho]\approx\frac{A'(0)}{B(0)}\rho=\frac{\frac{1-P}{P(1-K)^{2}}\sigma_{g}^{2}\varphi(t)^{2}}{\frac{K^{2}}{P^{2}}}\rho=\frac{P(1-P)}{K^{2}(1-K)^{2}}\sigma_{g}^{2}\varphi(t)^{2}\rho.\]

Hence, when the error of the approximation is small, the slope obtained
by regressing $Z_{ij}$ on $G_{ij}$ is an unbiased estimator of $\frac{P(1-P)}{K^{2}(1-K)^{2}}\sigma_{g}^{2}\varphi(t)^{2}$,
thus dividing it by $\frac{P(1-P)}{K^{2}(1-K)^{2}}\varphi(t)^{2}$
yields an unbiased estimator of $\sigma_{g}^{2}$ - the liability
scale heritability.

\subsection{Second order approximation}

While the first order approximation yields very satisfactory results
in our simulations, one can obtain a better estimator using a better
approximation. The second term of the Taylor series takes the form:

\[
\frac{A''(0)B(0)-2B'(0)A'(0)}{B(0)^{2}}\rho^{2}.\]

We have already derived:

\[
B(0)=\frac{K^{2}}{P^{2}},\]

and 

\[
A'(0)=\frac{1-P}{P(1-K)^{2}}\sigma_{g}^{2}\varphi(t)^{2}.\]

Now, $B(\rho)$ is the probability that two individuals with genetic
correlation $\rho$ are included in the study:

\[
B(\rho)=P(y_{i}=y_{j}=1;\rho,\sigma_{g}^{2})+\frac{K(1-P)}{P(1-K)}P(y_{i}\ne y_{j};\rho,\sigma_{g}^{2})+\Big(\frac{K(1-P)}{P(1-K)}\Big)^{2}P(y_{i}=y_{j}=0;\rho,\sigma_{g}^{2}).\]

Using the previous derivatives of the double integrals yields:

\[
B'(0)=\Big[1-2\frac{K(1-P)}{P(1-K)}+\Big(\frac{K(1-P)}{P(1-K)}\Big)^{2}\Big]\sigma_{g}^{2}\varphi(t)^{2}=\]

\[
\Big[1-\frac{K(1-P)}{P(1-K)}\Big]^{2}\sigma_{g}^{2}\varphi(t)^{2}=\Big[\frac{K-P}{P(1-K)}\Big]^{2}\sigma_{g}^{2}\varphi(t)^{2}.\]

Computing $A''(0)$ requires computing the second derivative of the
two-dimensional density at $\rho=0:$

\[
\int_{t}^{\infty}\int_{t}^{\infty}\frac{d^{2}}{d\rho^{2}}f_{\rho,\sigma_{g}^{2}}(l_{1},l_{2})=\int_{t}^{\infty}\int_{t}^{\infty}\sigma_{g}^{4}(l_{1}^{2}-1)(l_{2}^{2}-1)f_{0,\sigma_{g}^{2}}(l_{1},l_{2})=\sigma_{g}^{4}\varphi(t)^{2}t^{2},\]

so \[
A''(0)=\frac{1-P}{P(1-K)^{2}}\sigma_{g}^{4}\varphi(t)^{2}t^{2}.\]

Hence:

\[
\frac{A''(0)B(0)-2B'(0)A'(0)}{B(0)^{2}}=\frac{\frac{1-P}{P(1-K)^{2}}\sigma_{g}^{4}\varphi(t)^{2}t^{2}\frac{K^{2}}{P^{2}}-2\Big[\frac{K-P}{P(1-K)}\Big]^{2}\sigma_{g}^{2}\varphi(t)^{2}\frac{1-P}{P(1-K)^{2}}\sigma_{g}^{2}\varphi(t)^{2}}{\frac{K^{4}}{P^{4}}}\]

\[
=\frac{P}{K^{4}}\sigma_{g}^{4}\varphi(t)^{2}\frac{1-P}{(1-K)^{2}}\Big[t^{2}K^{2}-2\Big[\frac{K-P}{(1-K)}\Big]^{2}\varphi(t)^{2}\Big],\]

and the second order approximation can be written as:

\[
\mathbb{E}[Z_{ij}\mid s_{i}=s_{j}=1;\rho]\approx\frac{P(1-P)}{K^{2}(1-K)^{2}}\sigma_{g}^{2}\varphi(t)^{2}\rho\]

\[
+\frac{P}{K^{4}}\varphi(t)^{2}\frac{1-P}{(1-K)^{2}}\Big[t^{2}K^{2}-2\Big[\frac{K-P}{(1-K)}\Big]^{2}\varphi(t)^{2}\Big]\sigma_{g}^{4}\rho^{2}.\]

Since $K$ and $P$ are assumed to be known, the estimation problem
boils down to a single-variable non-linear regression in $\sigma_{g}^{2}$.

\section{Dealing with fixed effects }

\subsection{Extending the liability threshold model }

It is often desired to include fixed effects in the analysis of a
complex phenotype. Such fixed effects might include external information
such as sex, diet and exposure to environmental risks, but can also
be genetic variants with known effects or estimates of population
structure such as projections of several top principal components.

Since the liability threshold model is in fact a probit model, these
effects can be included in the usual manner:

\[
l_{i}=x_{i}^{\intercal}\beta+g_{i}+e_{i},\]

where $x_{i}$ is a vector of the values of the relevant covariates
and $\beta$ is a vector of their respective effect sizes. 

An individual is a case if $l_{i}>t$, as before. However, an equivalent
formulation would be to substract the fixed effects from the threshold,
rather than adding them to the liability:

\[
t_{i}=t-x_{i}^{\intercal}\beta,\]

thus keeping the previous formulation of the liability as a sum of
genetic and environmental effects.

\subsection{Heritability estimation with known fixed effects }

Assume the fixed effects are known, and so the $t_{i}$'s are known.
We define:

\[
K_{i}=P(y_{i}=1;t_{i}),\]

and:

\[
P_{i}=P(y_{i}=1\mid s_{i}=1;t_{i}),\]

to be the probability of the $i$'th individual being a case when
accounting for fixed effects through the adjustment of the threshold
$t_{i}$.

We redefine:

\[
Z_{ij}=\frac{(y_{i}-P_{i})(y_{j}-P_{j})}{\sqrt{P_{i}(1-P_{i})}\sqrt{P_{j}(1-P_{j})}},\]

so now $Z_{ij}$ can obtain four possible values:

\[
Z_{ij}=\begin{cases}
\frac{(1-P_{i})(1-P_{j})}{\sqrt{P_{i}(1-P_{i})}\sqrt{P_{j}(1-P_{j})}}\;\;\;\; & y_{i}=y_{j}=1\\
\frac{-P_{i}(1-P_{j})}{\sqrt{P_{i}(1-P_{i})}\sqrt{P_{j}(1-P_{j})}} & y_{i}=0,y_{j}=1\\
\frac{-P_{j}(1-P_{i})}{\sqrt{P_{i}(1-P_{i})}\sqrt{P_{j}(1-P_{j})}} & y_{i}=1,y_{j}=0\\
\frac{P_{i}P_{j}}{\sqrt{P_{i}(1-P_{i})}\sqrt{P_{j}(1-P_{j})}} & y_{i}=y_{j}=0\end{cases}.\]

We now wish to derive the same first order approximation before, while
conditioning on the fixed effects. 

Repeating the same steps as before while contitioning on $t_{i},t_{j}$
the expression for $A(\rho)$ is now:

\begin{eqnarray*}
A(\rho;t_i,t_j)&=&\frac{(1-P_{i})(1-P_{j})}{\sqrt{P_{i}(1-P_{i})}\sqrt{P_{j}(1-P_{j})}}\mathbb{P}(y_{i}=y_{j}=1;\rho,t_{i},t_{j})+\\
&&\frac{K(1-P)}{P(1-K)}\frac{-P_{i}(1-P_{j})}{\sqrt{P_{i}(1-P_{i})}\sqrt{P_{j}(1-P_{j})}}\mathbb{P}(y_{i}=0,y_{j}=1;\rho,t_{i},t_{j})+\\
&&\frac{K(1-P)}{P(1-K)}\frac{-P_{j}(1-P_{i})}{\sqrt{P_{i}(1-P_{i})}\sqrt{P_{j}(1-P_{j})}}\mathbb{P}(y_{i}=1,y_{j}=0;\rho,t_{i},t_{j})+\\
&&\Big(\frac{K(1-P)}{P(1-K)}\Big)^{2}\frac{P_{i}P_{j}}{\sqrt{P_{i}(1-P_{i})}\sqrt{P_{j}(1-P_{j})}}\mathbb{P}(y_{i}=0,y_{j}=0;\rho,t_{i},t_{j}).
\end{eqnarray*}

Each phenotype was standardized to have mean $0$, conditional on
the relevant fixed effects. Additionaly, with $\rho=0$ the phenotypes
are independet and so the expected value of $Z$ at $\rho=0$ is $0$.
An immediate result is that $A(0)=0$. To see this consider the Taylor
expansion of 

\[
\mathbb{E}[Z_{ij}\mid s_{i}=s_{j}=1;\rho,t_{i},t_{j}]=\frac{A(0;t_{i},t_{j})}{B(0;t_{i},t_{j})}+\sum_{i=1}^{\infty}c_{i}\rho^{i},\]

for some constants $\{c_{i}\}_{i=1}^{\infty}$. Setting $\rho=0$
yields:

\[
\mathbb{E}[Z_{ij}\mid s_{i}=s_{j}=1;0,t_{i},t_{j}]=\frac{A(0;t_{i},t_{j})}{B(0;t_{i},t_{j})},\]

but on the other hand:

\[
\mathbb{E}[Z_{ij}\mid s_{i}=s_{j}=1;0,t_{i},t_{j}]=\mathbb{E}\Big[\frac{(y_{i}-P_{i})(y_{j}-P_{j})}{\sqrt{P_{i}(1-P_{i})}\sqrt{P_{j}(1-P_{j})}}\Big]=\]

\[
\mathbb{E}\Big[\frac{y_{i}-P_{i}}{\sqrt{P_{i}(1-P_{i})}}\Big]\mathbb{E}\Big[\frac{y_{j}-P_{j}}{\sqrt{P_{j}(1-P_{j})}}\Big]=0.\]
Therefore, $\frac{A(0;t_{i},t_{j})}{B(0;t_{i},t_{j})}=0$ and so $A(0;t_{i},t_{j})=0$. We conclude that the first order Taylor approximation is again of the form:
\[
\mathbb{E}[Z_{ij}\mid s_{i}=s_{j}=1;\rho,t_{i},t_{j}]\approx\frac{A'(0;t_{i},t_{j})}{B(0;t_{i},t_{j})}\rho.\]

To derive an explicit expression for $A'(0;t_{i},t_{j})$ we need
to differentiate the double integrals as before. When both $i$ and $j$ are cases, with thresholds $t_{i},t_{j}$ respectively, the double integral takes
the form:

\[\int_{t_{i}}^{\infty}\int_{t_{j}}^{\infty}f_{\rho,\sigma_{g}^{2}}(l_{1},l_{2})dl_{1}dl_{2},\]
differentiating it w.r.t. $\rho$ and setting $\rho=0$ yields:
\[
\varphi(t_{i})\varphi(t_{j})\sigma_{g}^{2},\]
and differentiating the other double integrals yields similar results:
\[
\frac{d}{d\rho}\int_{-\infty}^{t_{i}}\int_{-\infty}^{t_{j}}f_{\rho,\sigma_{g}^{2}}(l_{1},l_{2})dl_{1}dl_{2}\mid_{\rho=0}=\varphi(t_{i})\varphi(t_{j})\sigma_{g}^{2},
\]
and:
\begin{eqnarray*}
\frac{d}{d\rho}\int_{t_{i}}^{\infty}\int_{-\infty}^{t_{j}}f_{\rho,\sigma_{g}^{2}}(l_{1},l_{2})dl_{1}dl_{2}\mid_{\rho=0}&=&\\
\frac{d}{d\rho}\int_{t_{j}}^{\infty}\int_{-\infty}^{t_{i}}f_{\rho,\sigma_{g}^{2}}(l_{1},l_{2})dl_{1}dl_{2}\mid_{\rho=0}&=&-\varphi(t_{i})\varphi(t_{j})\sigma_{g}^{2},
\end{eqnarray*}

We differentiate $A(\rho;t_{i},t_{j})$ w.r.t. $\rho$, and using the previous results and some algebra we get:

\[A'(0;t_{i},t_{j})=\frac{\varphi(t_{i})\varphi(t_{j})\sigma_{g}^{2}}{\sqrt{P_{i}(1-P_{i})}\sqrt{P_{j}(1-P_{j})}}\Big[1-(P_{i}+P_{j})(1-\frac{K(1-P)}{P(1-K)})+P_{i}P_{j}\Big(1-\frac{K(1-P)}{P(1-K)}\Big)^{2}\Big]=\]

\[\frac{\varphi(t_{i})\varphi(t_{j})\sigma_{g}^{2}}{\sqrt{P_{i}(1-P_{i})}\sqrt{P_{j}(1-P_{j})}}\Big[1-(P_{i}+P_{j})\Big(\frac{P-K}{P(1-K)}\Big)+P_{i}P_{j}\Big(\frac{P-K}{P(1-K)}\Big)^{2}\Big].\]

As a sanity check, set $P_{i}=P_{j}=P$ and $t_{i}=t_{j}=t$ to get:
\[
\frac{\varphi(t)^{2}\sigma_{g}^{2}}{P(1-P)}\Big[1-2P\Big(\frac{P-K}{P(1-K)}\Big)+P^{2}\Big(\frac{P-K}{P(1-K)}\Big)^{2}\Big]=\frac{\varphi(t)^{2}\sigma_{g}^{2}}{P(1-P)}\Big(\frac{1-P}{1-K}\Big)^{2}=\frac{\varphi(t)^{2}\sigma_{g}^{2}(1-P)}{P(1-K)^{2}},\]
which is the expression derived for the no fixed effects case.

Moreover, the the probability of inclusion of individual $i$ in the
study, conditional on the fixed effects is now:

\[
K_{i}+(1-K_{i})\frac{K(1-P)}{P(1-K)}.\]
and so:
\[
B(0;t_{i},t_{j})=\Big(K_{i}+(1-K_{i})\frac{K(1-P)}{P(1-K)}\Big)\Big(K_{j}+(1-K_{j})\frac{K(1-P)}{P(1-K)}\Big).\]

Plugging the derived expressions for $A'(0;t_{i},t_{j}),B(0;t_{i},t_{j})$,
we conclude that regressing $Z_{ij}$ on 

\[
\frac{\varphi(t_{i})\varphi(t_{j})\Big[1-(P_{i}+P_{j})\Big(\frac{P-K}{P(1-K)}\Big)+P_{i}P_{j}\Big(\frac{P-K}{P(1-K)}\Big)^{2}\Big]}{\sqrt{P_{i}(1-P_{i})}\sqrt{P_{j}(1-P_{j})}\Big(K_{i}+(1-K_{i})\frac{K(1-P)}{P(1-K)}\Big)\Big(K_{j}+(1-K_{j})\frac{K(1-P)}{P(1-K)}\Big)}G_{ij},\]
yields an estimator of heritability on the liability scale.

\subsection{Dealing with unknown fixed effects }

More often than not, the effects of relevant fixed effects are unknown
and must be estimated from the data. However, estimating effect sizes
under ascertainment in case-control studies is notoriously problematic.
Specifically, under the threshold (probit) model, ignoring the ascertainment
yields biased estimators.

A special exception is the case of logistic regression. In their seminal
paper, Prentice and Pyke (1979)\citep{prentice1979logistic} proved that using a logistic regression to
estimate fixed effects from ascertained data yields consistent estimators
of these effects in the (unascertained) population, and that the ascertainment
only biases the intercept. 

We therefore suggest a two-step procedure for estimating heritability.
First, we estimate the fixed effects using a logistic regression model.
We then correct the effect of the ascertainment, and obtain the individual-specific
thresholds. Lastly, we plug the thresholds into the estimation scheme
described above. 

More elaborately, by Bayes' formula:
\[P_{i}=\frac{\mathbb{P}(s_{i}=1\mid y_{i}=1;x_{i})K_i}{\mathbb{P}(s_{i}=1 ; x_{i})},\]
by the complete ascertainment assumption $\mathbb{P}(s_{i}=1\mid y_{i}=1,x_{i})=1$,
and according to the selection scheme:
\[\mathbb{P}(s_{i}=1 ; x_{i})=K_{i}+\frac{K(1-P)}{P(1-K)}\big(1-K_{i}\big).\]
We can thus solve for $K_i$ and express it is a function of $P_i$:
\[
K_{i} =  \frac{\frac{K(1-P)}{P(1-K)} P_{i}}{1 + \frac{K(1-P)}{P(1-K)} P_{i} - P_{i}}  
\]        

We then use logistic regression to obtain  $\hat{P}_{i}$ - a consistent estimator of $P_{i}$, and use this estimate to obtain an estimate of $K_i$, which is in turn used to estimate the threshold: 
\[\hat{t}_{i}=\Phi^{-1}\big(1-\hat{K}_{i}\big),\]
and the estimates of the individual-wise thresholds are used for estimating
the liability-scale heritability.

\subsubsection{Estimating the added variance due to fixed effects}
Lastly, the presence of fixed effects increases the variance of the liability, so $\sigma_{g}^{2}$ no longer equals $h^2$. The appropriate definition of heritability is now: 
\[
h^2=\frac{\sigma_{g}^{2}}{\sigma_{g}^{2}+\sigma_{e}^{2}+\sigma_{t}^{2}}=\frac{\sigma_{g}^{2}}{1+\sigma_{t}^{2}},
\]
where $\sigma_{t}^{2}$ is the variance of the thresholds in the population, and so  the estimate of $\sigma_{g}^{2}$ can be transformed to an estimate of the heritability simply by dividing it by  $1+\sigma_{t}^{2}$. Therefore, obtaining an estimate of the heritability in the presence of fixed-effects requires an estimate of $\sigma_{t}^{2}$ in the population. 
To estimate $\sigma_{t}^{2}$ we use the law of total variance:
\[
\sigma_{t}^{2}=V(t)=V(\mathbb{E}(t \mid y))+\mathbb{E}(V(t \mid y))
\]
where y is the phenotype of the individual. Furthermore:
\[
V(\mathbb{E}(t \mid y))=K(1-K)(\mathbb{E}(t \mid y=1)-\mathbb{E}(t \mid y=0))^2,
\]
and:
\[
\mathbb{E}(V(t \mid y))=KV(t \mid y=1) + (1-K)V(t \mid y=0).
\]
Once we condition on a specific phenotype, both the expected and variance of $T$ can be estimated from the data, as they are no longer affected by enrichment of cases in the data. Specifically, we use the estimated thresholds $\hat{t}_{i}$ to estimate the expected value and the variance of the threshold for cases and controls, and plug the estimates into the equations above to obtain $\hat{\sigma}_{t}^{2}$.

We discuss how $\sigma_{t}^{2}$ can be estimated from the data in the Supplementary material.  

\section{Simulations }

\subsection{Description of the Lee et al. simulation scheme}

Lee et al. perform simulations of ascertained case-control studies
in the following manner: 

Phenotypes are simulated in blocks of $100$ individuals Given the
true genetic variance $\sigma_{g}^{2}$, liabilities for each are
sampled from a multivariate Gaussian distribution with mean $0$ and
covariance matrix $\Sigma=G\sigma_{g}^{2}+I(1-\sigma_{g}^{2})$, where
$G$ is given by:

\[
G_{ij}=\begin{cases}
1 & i=j\\
0.05 & i\ne j\end{cases},\]

that is -- individuals within each block have a genetic correlation
of $0.05$ while individuals in different blocks are perfectly unrelated.
Individuals with liabilities higher than the threshold $t$ are considered
cases.

Cases are always included in the study. Controls are included in the
study with probability $\frac{K}{1-K}$. Blocks are generated until
$100$ individuals are accumulated, and this process is repeated $100$
times, so in total $10,000$ cases and controls are accumulated. 


\subsection{Discussion of problems with the simulations in Lee et al.}

There are several key aspects in which the simulations of Lee et al.
differ from the true generative process of the data.

First, the resulting correlation matrix is highly degenerate, with
most of the correlations being 0 -- as much as 99.99\textbackslash{}\%.
Since the underlying idea behind heritability estimation from unrelated
individuals is leveraging the minor -- but non-zero -- correlations
among unrelated individuals, restricting the simulations to largely
degenerate correlation structures is highly unrealistic and counterproductive. 

Second, in reality, correlations between individuals span a wide range
of values. while in Lee et al.'s simulations the correlations are
either $0$ or $0.05.$ Hence negative correlations are impossible,
and the expected correlation is larger than $0$, contrary to the
unrelated individuals assumption.

Thirdly, the selection procedure, in which cases are up-sampled compared
to their prevalence in the population, results in cases being more
similar genetically than controls. This is a well known phenomenon
which has recently attracted considerable attention in the context
of GWAS \cite{zaitlen2012analysis}. This is hardly captured by the simulation
process of Lee et al. For example, cases from different blocks always
have 0 genetic correlation between them in Lee et al.'s simulations,
while in a realistic scenario they should have positive genetic correlation,
whose magnitude depends on the heritability of the phenotype and the
number of causative loci. 

Lastly, for small enough prevalence, all blocks in a typical simulation
would contain only a single individual. In this case the genetic effects
would be completely interchangeable with the environmental effects
since $G=I$ and so heritability would be impossible to estimate.
In reality, however, a very small prevalence should yield very closely
related individuals, thus generating an intuitively easier estimation
problem.

We also note that Lee et al.'s simulations use an unrealistic number
of individuals - 10,000.

\subsection{Simulations using a generative model}

\begin{figure}[H]
\includegraphics[scale=0.7]{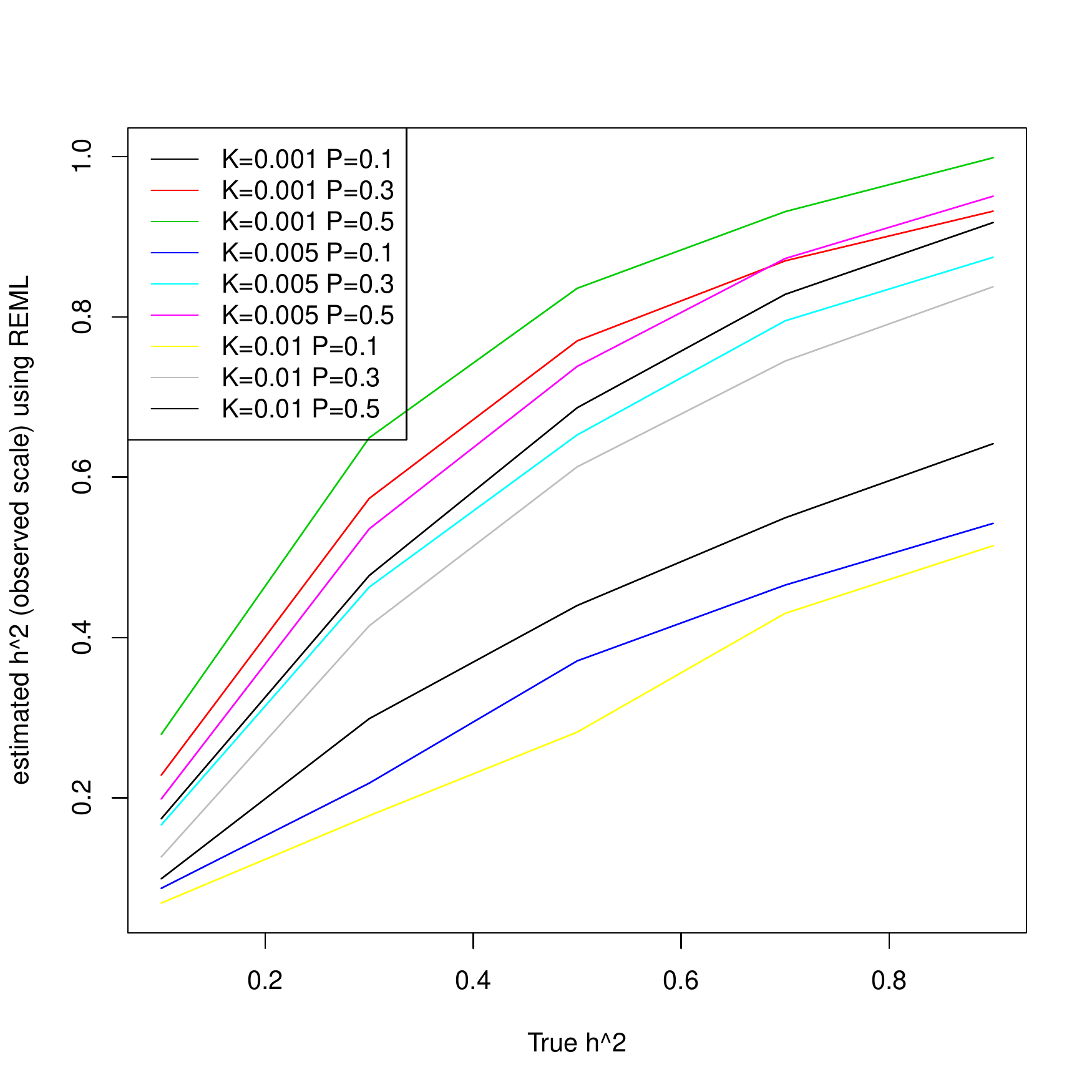}

\caption{Estimated heritability on the observed scale for various values of
$\sigma_{g}^{2},K$ and $P$, demonstrating the non-linearity of $\hat{h_{o}^{2}}$
(as a function of $\sigma_{g}^{2}$). }
\label{Flo:REML_obs}
\end{figure}

To see if these problems have a major impact, we ran simulations using
the full generative model. This was done as follows:
\begin{enumerate}
\item The MAFs of 10,000 MAFs were randomly sampled from $U[0.05,0.5]$. 
\item SNP effect sizes were randomly sampled from $N(0,\frac{\sigma_{g}^{2}}{m}).$
\item For each individual, we:

\begin{enumerate}
\item Randomly generated a genotype using the MAFs, and normalized it (according
to Yang's model).
\item Used the genotype and the effect sizes, to compute the genetic effect.
\item Sampled an environmental effect from $N(0,1-\sigma_{g}^{2})$.
\item Computed liability and phenotype.
\item If the phenotype was a case - the individual was automatically included
in the study. Otherwise the individual was included in the study with
probability $\frac{K(1-P)}{P(1-K)}$. 
\end{enumerate}
\item Step (2) was repeated until enough individuals were accumulated (4,000). 
\item The genotypes of all included individuals were used to compute $G=\frac{ZZ^{\intercal}}{m}$
where $Z$ is the matrix of normalized genotypes. 
\item We used GCTA \citep{yang2011gcta} to estimate heritability with $G$ as the genetic
correlation matrix, and a $0-1$ vector of the phenotypes as input. 
\end{enumerate}
We ran ten repetitions of this simulation for all combinations of
$\sigma_{g}^{2}\in\{0.1,0.3,0.5,0.7,0.9\}$, $P\in\{0.1,0.3,0.5\}$
and $K\in\{0.001,0.005,0.01\}$. The results for all combinations
of are given in figure \ref{Flo:REML_obs}. Note how the estimated heritability on the
observed scale is not linear in the true underlying heritability.
Moreover, the simulated sets with less ascertainment are more linear
(most notably $P=0.1$ and $K=0.01$ - the yellow line in figure
\ref{Flo:REML_obs}).

\begin{figure}[H]
\includegraphics[scale=0.7]{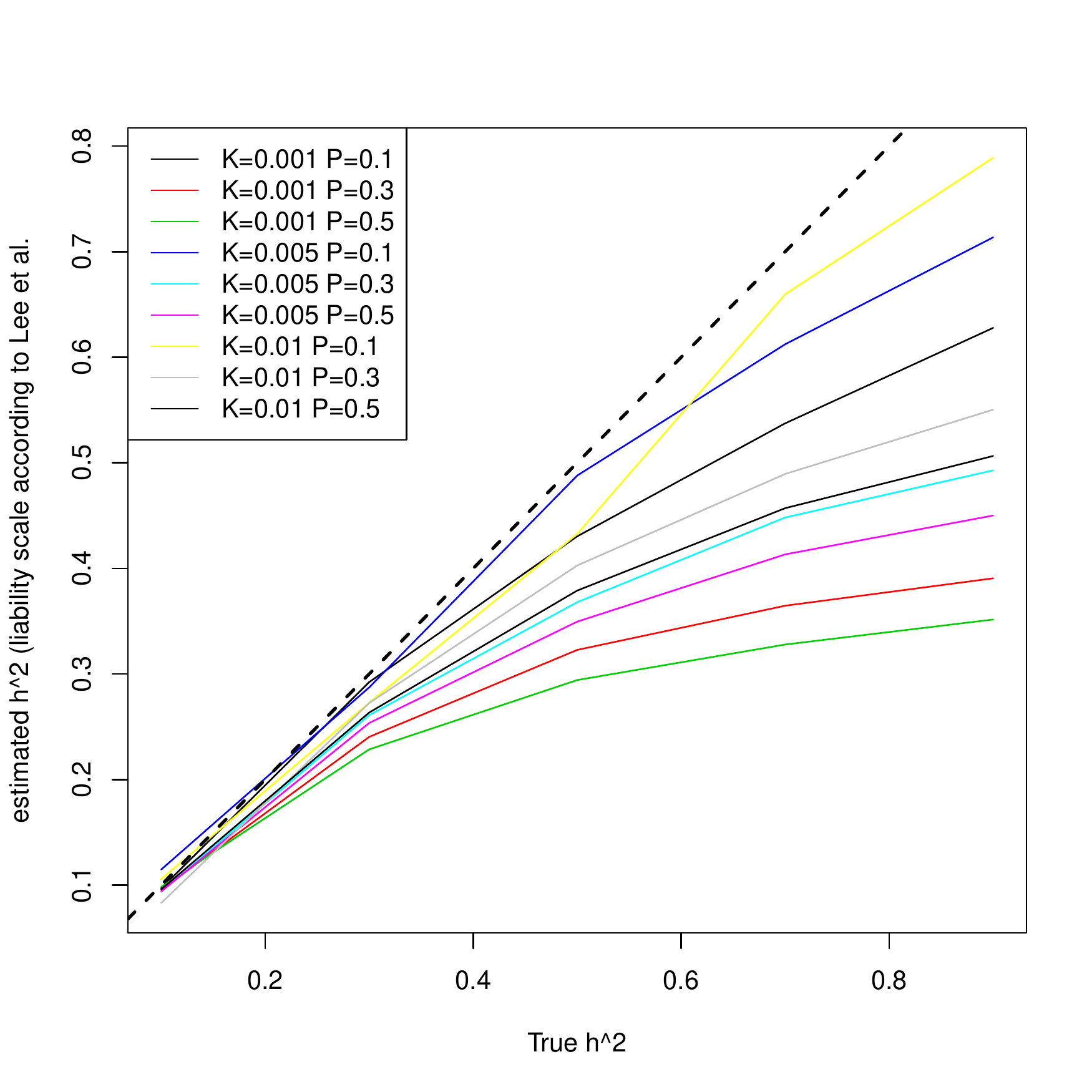}

\caption{Estimated heritability on the observed scale for various values of
$\sigma_{g}^{2},K$ and $P$, demonstrating the non-linearity of $\hat{h_{o}^{2}}$
(as a function of $\sigma_{g}^{2}$), even after correcting the estimate
as per Lee et al. The dashed line is $y=x$, i.e. the true heritability. }
\label{Flo:REML_cor}
\end{figure}

Of course, since the correction suggested in Lee et al. is linear
and independent of the true underlying heritability, applying it to
the data does not yield accurate estimates, as can be seen in figure \ref{Flo:REML_cor}.

\begin{figure}[H]
\includegraphics[scale=0.7]{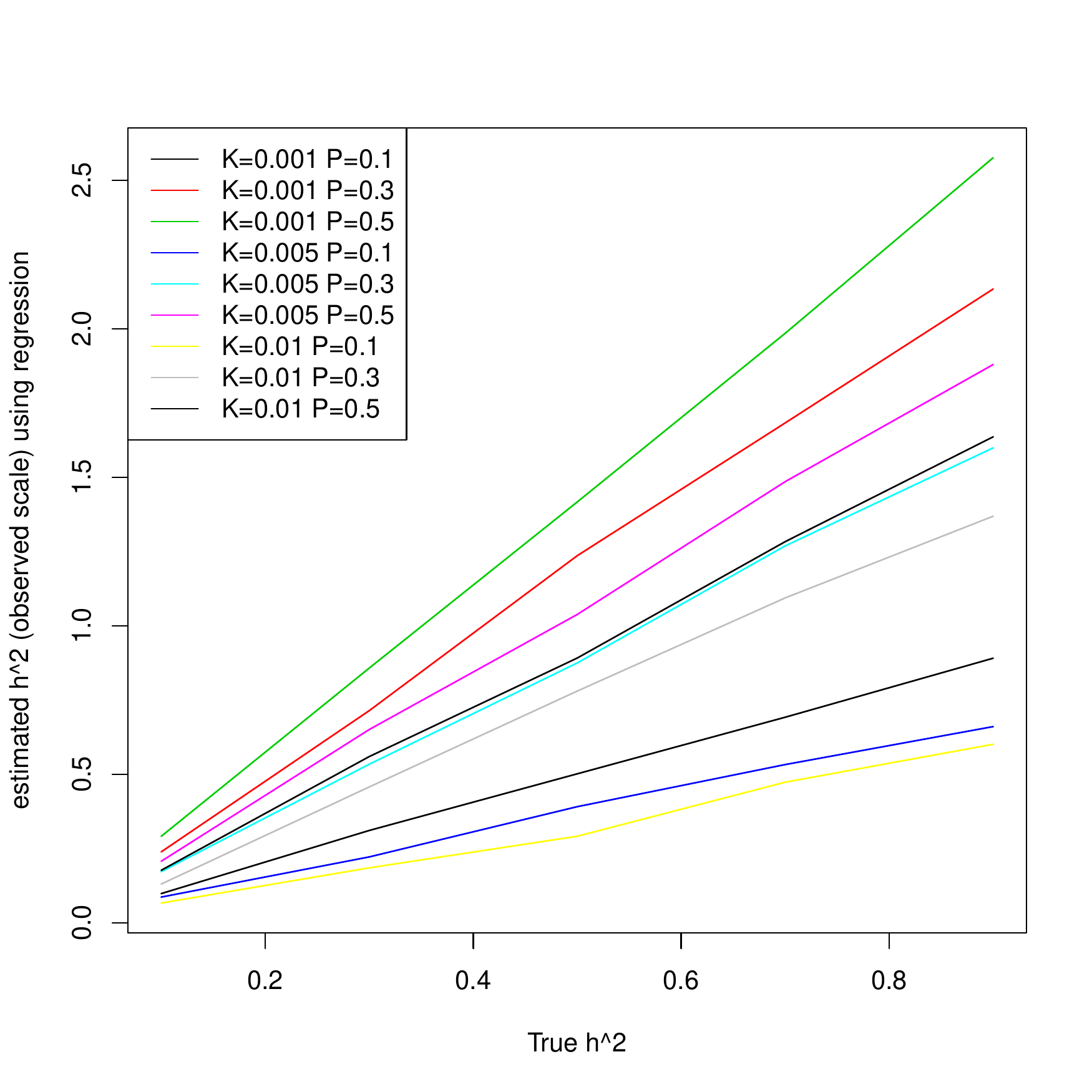}

\caption{Average estimates obtained by GCR using the same simulations as in figures 1-2, before correction.}
\label{Flo:reg_obs}
\end{figure}

\begin{figure}[H]
\includegraphics[scale=0.7]{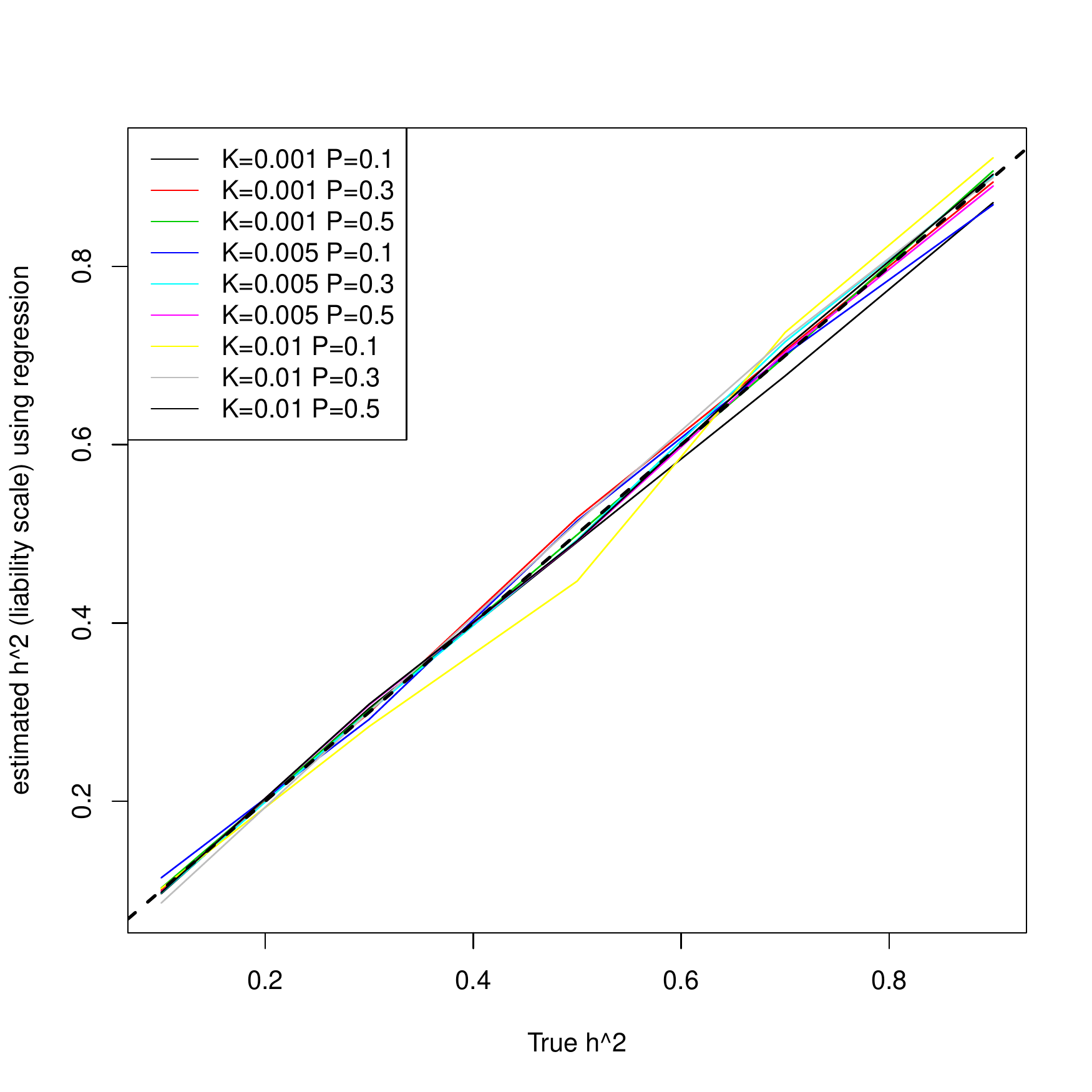}

\caption{Corrected average estimates obtained by GCR using the same simulations as in figures 1-2. The dashed line is $y=x$, indicating the estimators are indeed unbiased.}
\label{Flo:reg_cor}
\end{figure}

We then applied GCR to the same data. The results are portrayed
in figure \ref{Flo:reg_obs}. As can clearly be seen from the figure,
our estimators are linear in the underlying true heritability. Indeed,
as can be seen in \cref{Flo:reg_cor,Flo:reg_cor_box}, applying the correction
derived earlier yielded accurate estimators. 

\subsubsection{Distribution of estimates around the mean }

\begin{figure}[H]
\includegraphics[scale=0.7]{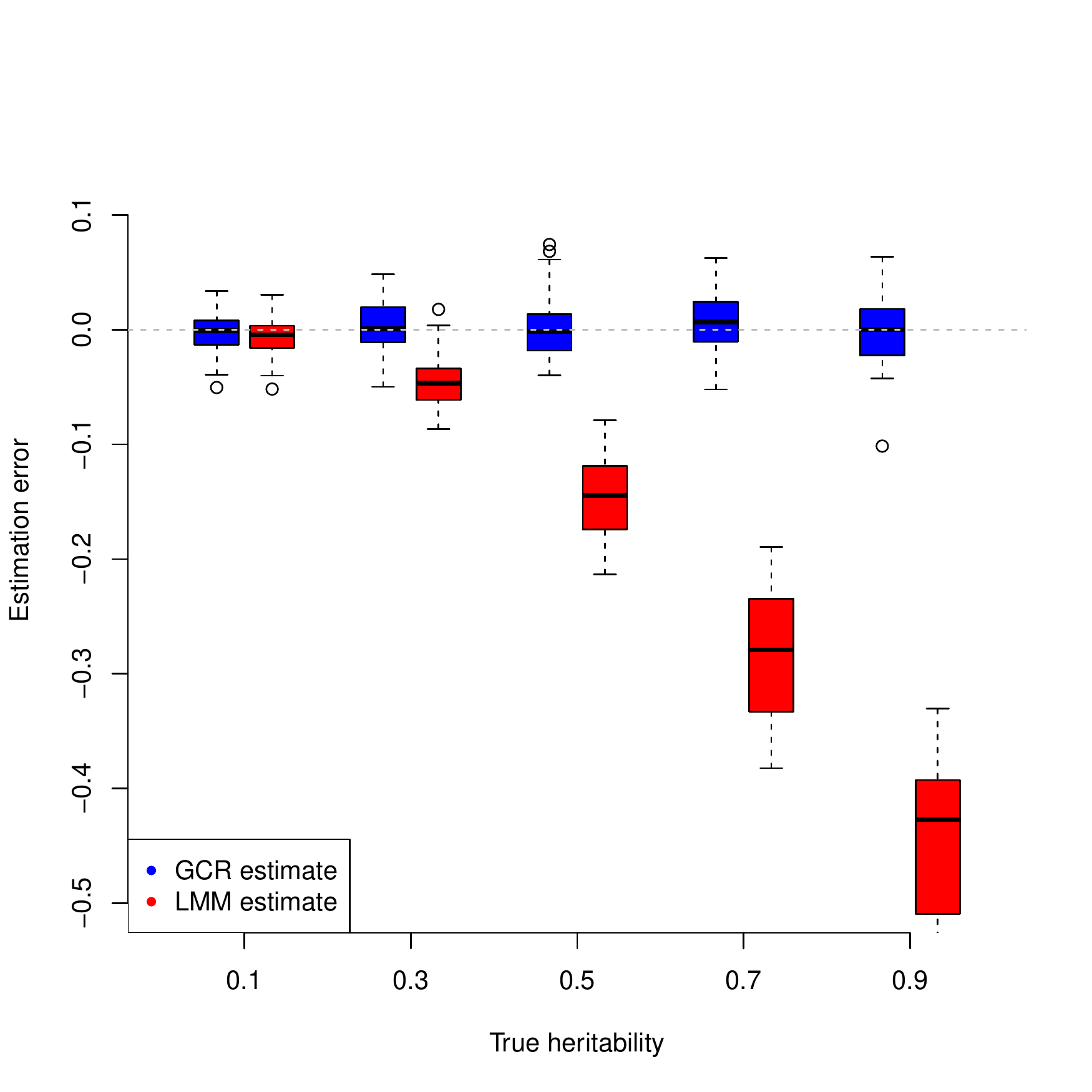}

x
\caption{Distribution of estimation errors for various values of $h^{2}$.
Estimates for different values of $K$ and $P$ are grouped together. }
\label{Flo:reg_cor_box}
\end{figure}

Apart from being unbiased, GCR estimates display an equal or lower variance to the LMM estimates. This can be seen in figure \ref{Flo:reg_cor_box}.

\subsection{Additional simulations }

\subsubsection{Weak ascertainment}
\begin{figure}[H]
\includegraphics[scale=0.7]{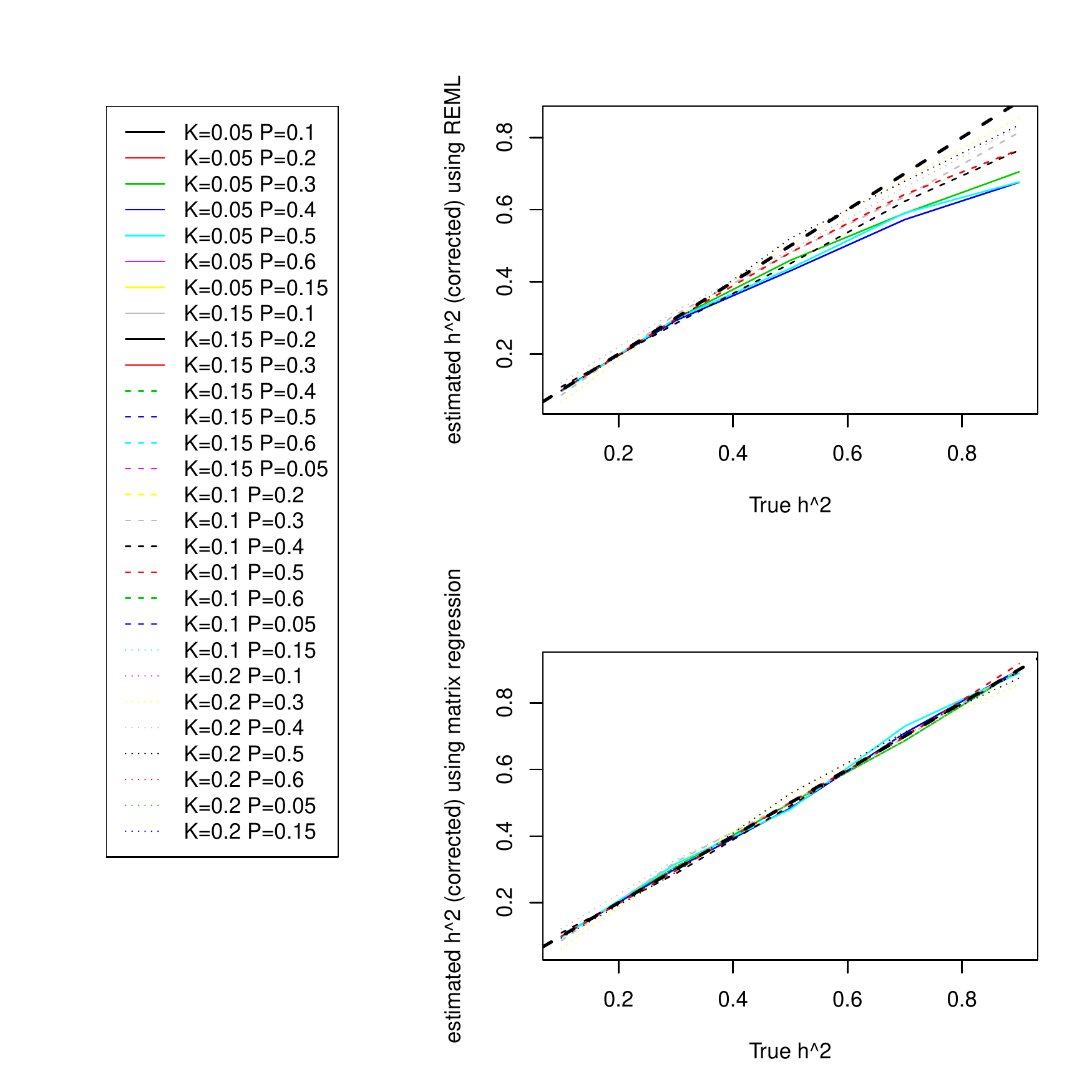}

\caption{Comparing corrected LMM estimates (top) to GCR estimates
(bottom) for studies with low-to-intermediate ascertainment.}
\label{Flo:low_asc}
\end{figure}

To test the robustness of our method to the different values of $P$
and $K$, we ran a wide range of simulations, with less extreme ascertainment.
The results are given in figure \ref{Flo:low_asc}, showing how the
downward-bias of the LMM estimates is smaller for less ascertained studies, and disappears completely when the studies are non-ascertained.
in both cases the GCR yields unbiased estimates.

\subsubsection{non-ascertained simulations}

\begin{figure}[H]
\includegraphics[scale=0.7]{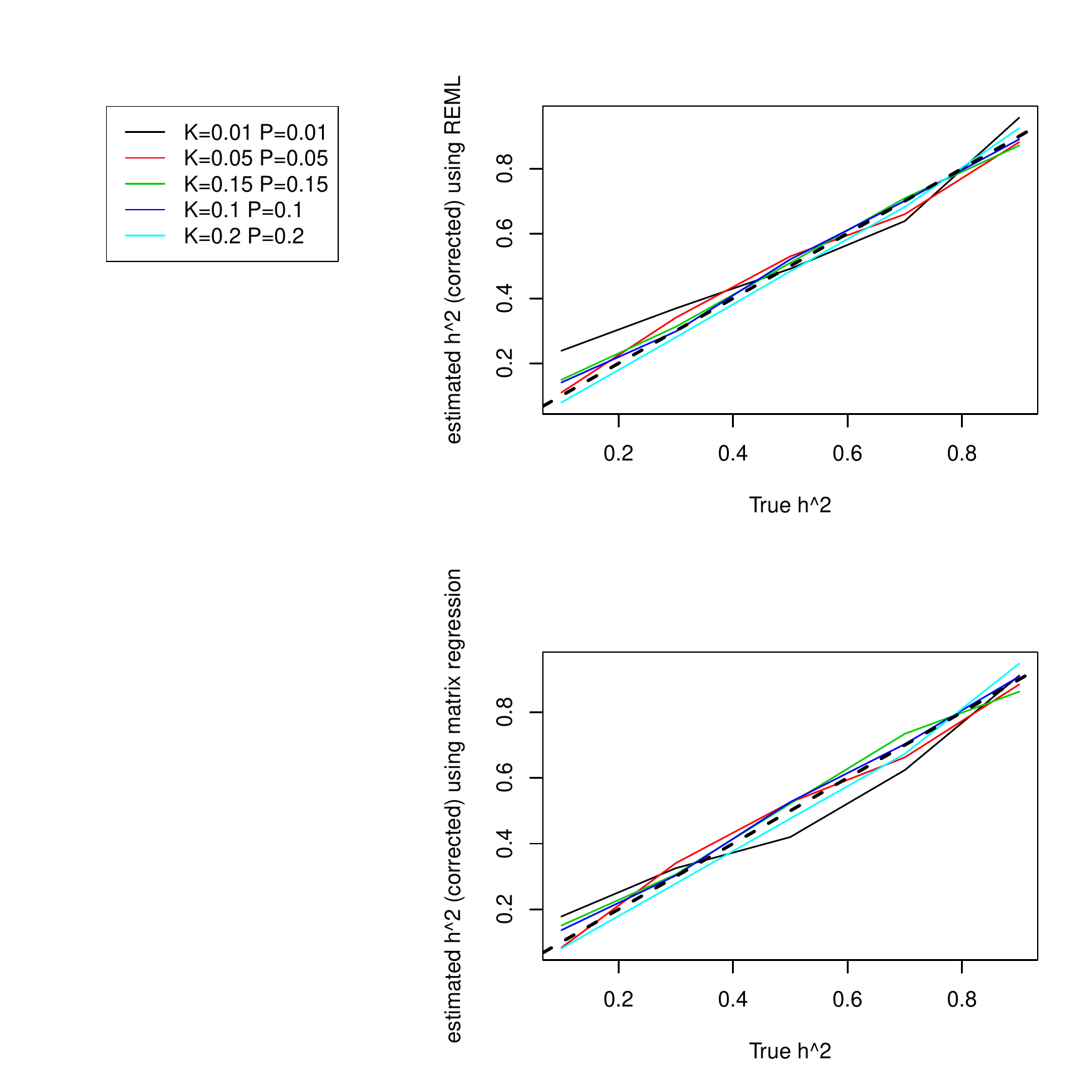}

\caption{Comparing corrected LMM estimates (top) to GCR estimates
(bottom) for studies with no ascertainment. Both methods yield unbiased
estimates, except for low values of heritability, where the positivity constraint introduces a mild positive bias.}
\label{Flo:no_asc}
\end{figure}

We also applied our method to non-ascertained simulations, where both the
LMM and GCR methods produce unbiased estimators, as can be seen in figure
\ref{Flo:no_asc}. 

\subsubsection{Simulations with population structure}

\begin{figure}[H]
\includegraphics[scale=0.7]{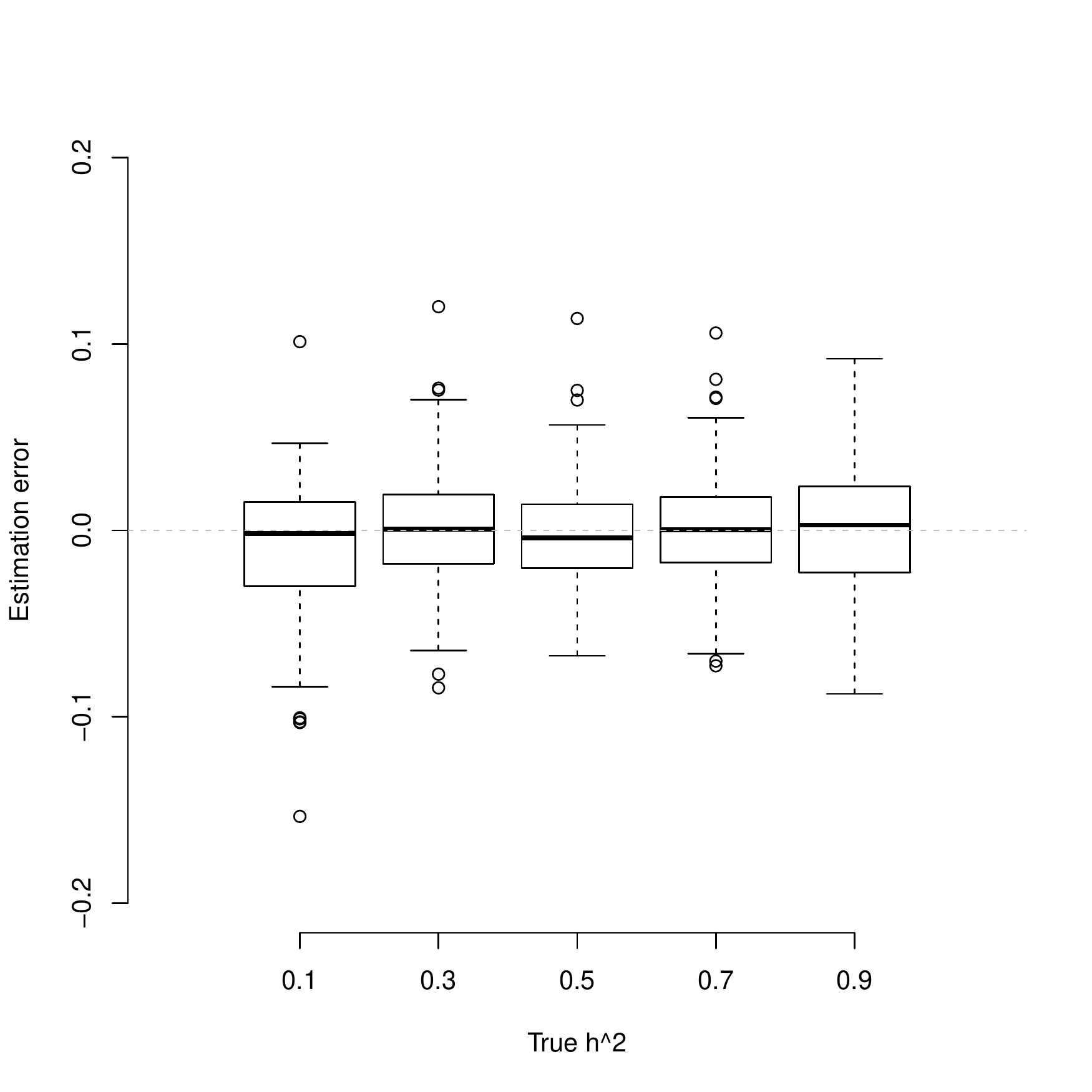}

\caption{Estimation errors for simulations generated with mild population structure. GCR estimates are still unbiased in the presence of such structure.}
\label{Flo:pop_structure}
\end{figure}

To study the robustness of our method to the presence of subtle population structure, we modified our simulations to introduce population structure similar to the simulations in Lee et al. 
For every batch of 100 individuals, we randomly selected 5\% of the SNPs and fix them for the entire batch. By doing so, the expected correlation within each batch is $0.05$, as in Lee et al. while the actual realized correlations are still realistic. The process of phenotype generation and individual selection then proceeds as described earlier. We ran simulations with $\sigma_{g}^{2}\in\{0.1,0.3,0.5,0.7,0.9\}$, $P\in\{0.1,0.3,0.5\}$
and $K\in\{0.001,0.005,0.01\}$. The resulting GCR estimates were unbiased for all these scenarios, as can be seen in figure \ref{Flo:pop_structure}.

\subsubsection{Simulations with different numbers of SNPs}

\begin{figure}[H]
\includegraphics[scale=0.7]{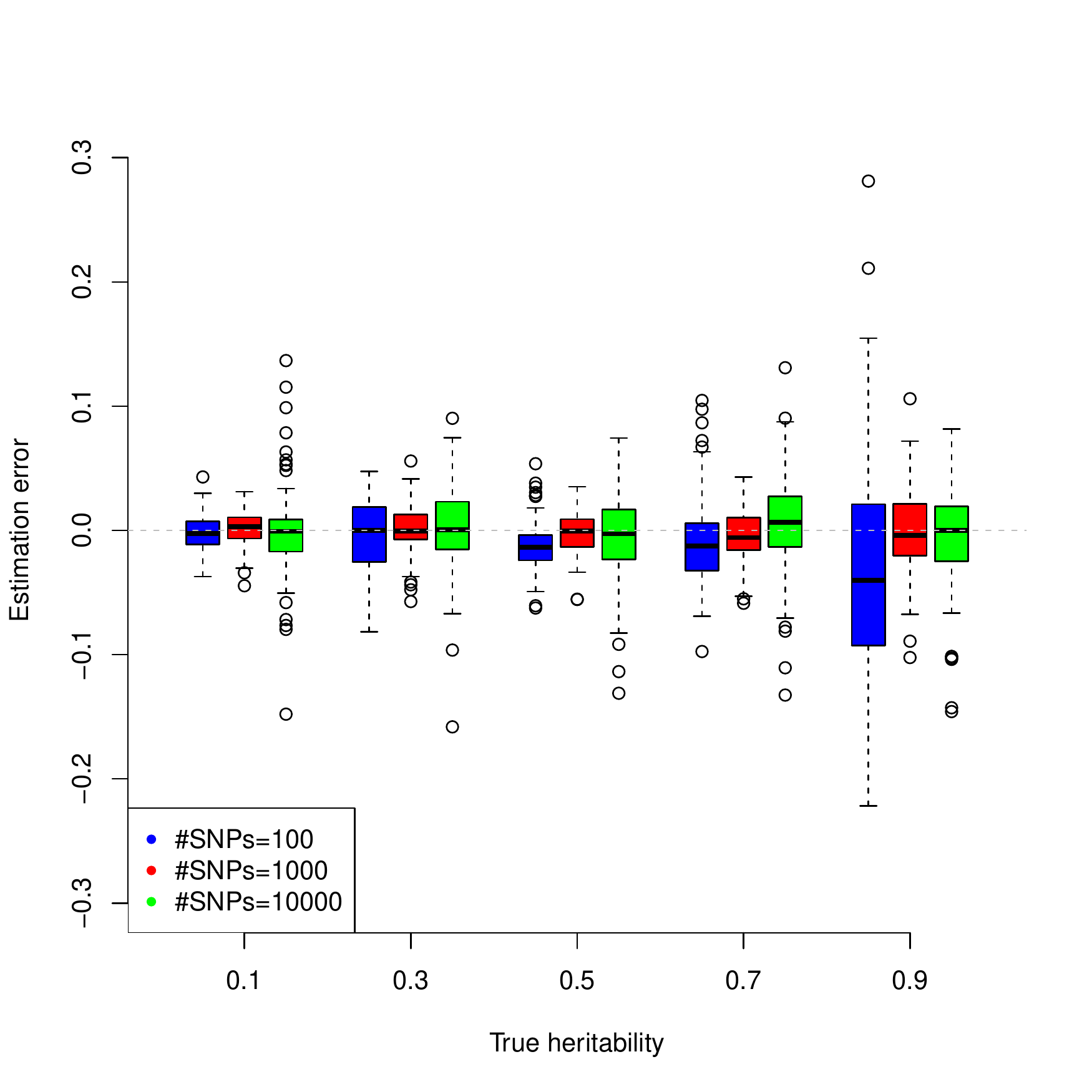}

\caption{Comparing GCR estimates for various simulation scenarios with different number of SNPs used to simulate the phentoypes and compute the genetic correlations.}
\label{Flo:snp_numbers}
\end{figure}

The number of SNPs used in the simulations determines the variance of the realized genetic correlations between individuals. Since GCR relies on a first-order apporximation around zero correlation, high variance of genetic correlations might decrease the accuracy of GCR. To study the robustness of GCR to such changes, we reran our initial simulations using 1,000 or 100 SNPs. GCR still produced unbiased estimates. As expected, when the number of SNPs was smallest (100) and the heritability was highest (0.9), the results displayed cosiderably higher variance. The results are summarized in figure \ref{Flo:snp_numbers}

\subsubsection{Second order simulations}

\begin{figure}[H]
\begin{minipage}[t]{0.45\columnwidth}%
\includegraphics[scale=0.45]{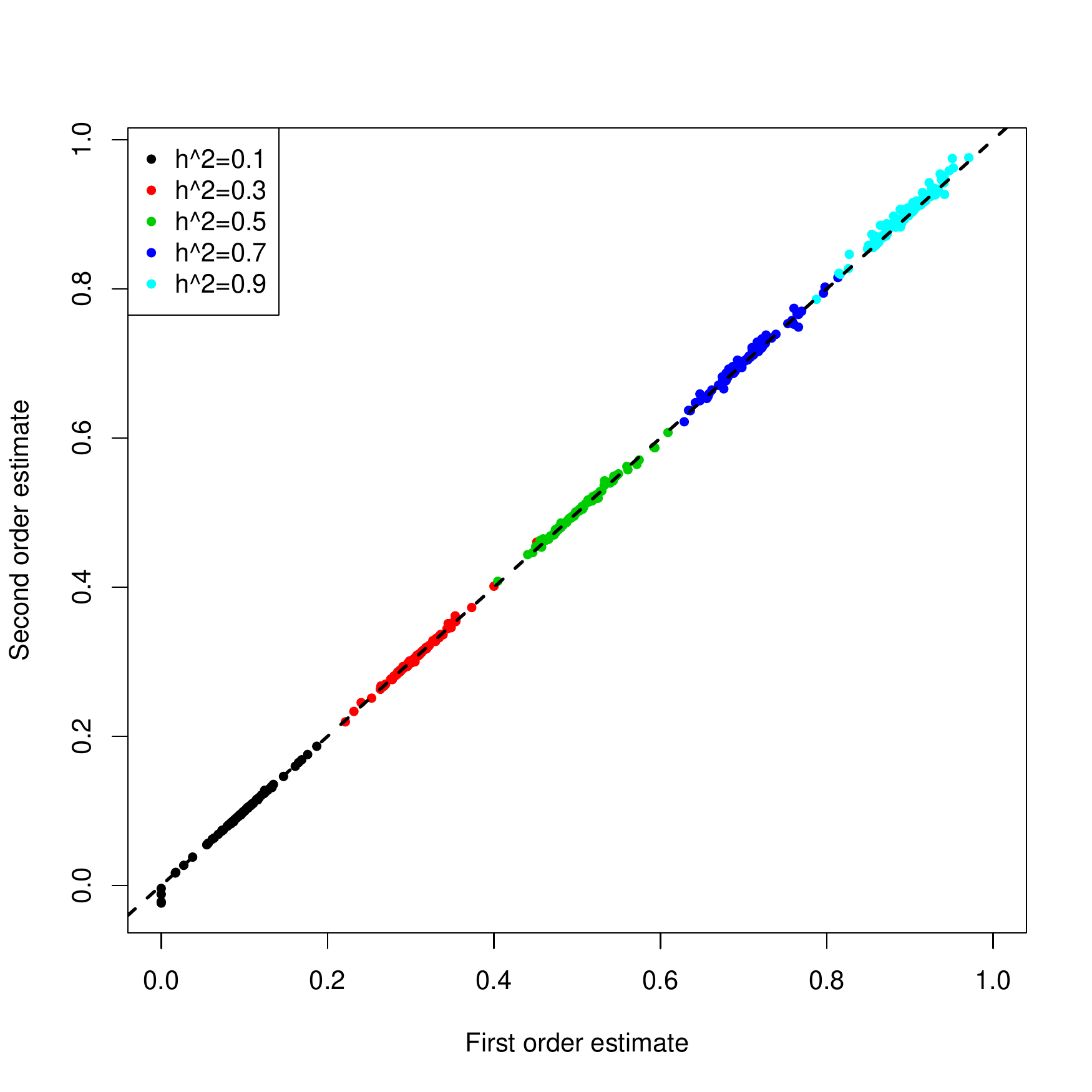}%
\end{minipage}\hfill{}%
\begin{minipage}[t]{0.45\columnwidth}%
\includegraphics[scale=0.45]{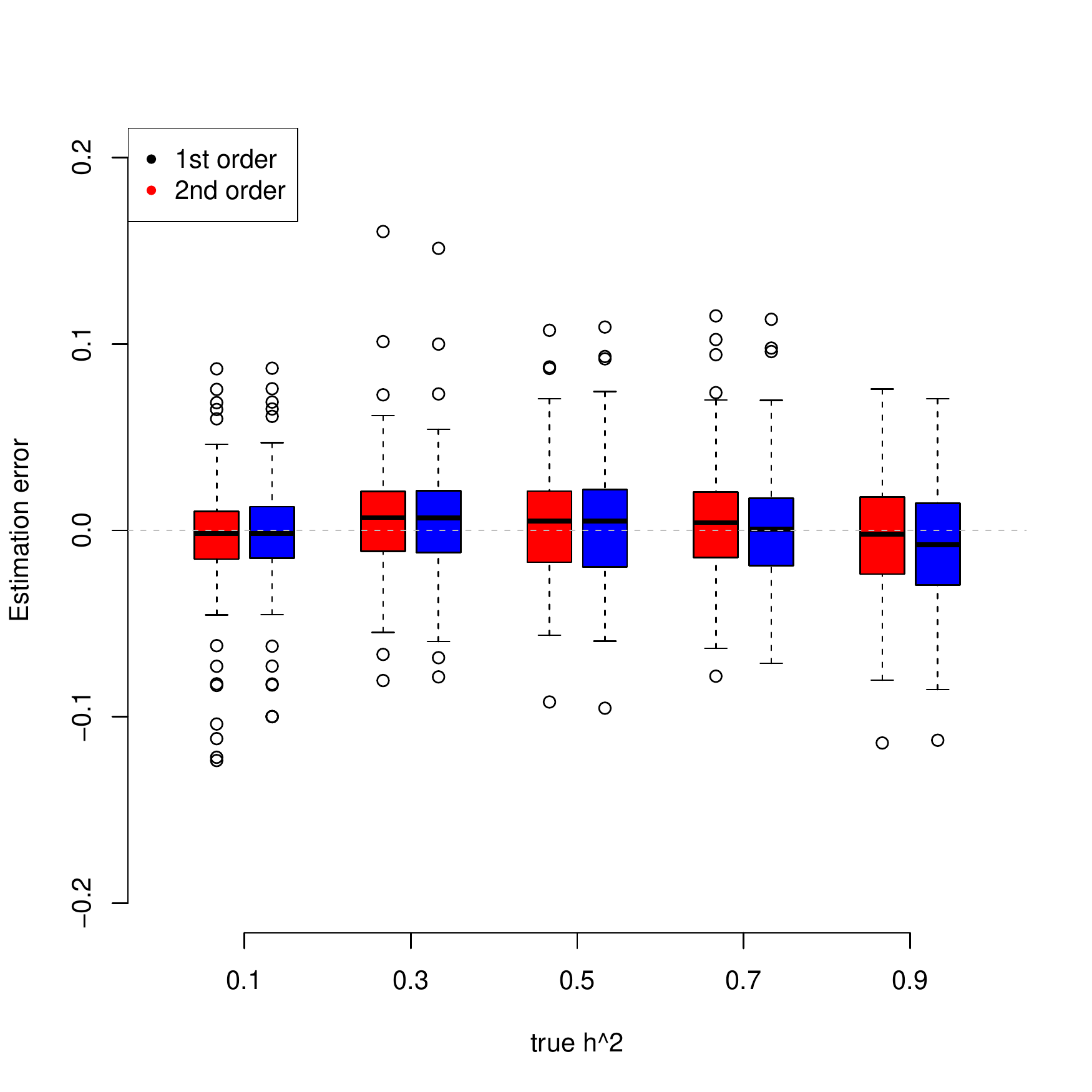}%
\end{minipage}
\caption{Comparing the first and second order GCR heritability estimates. Left: Plotting the 1st vs. 2nd order estimates for each simulation run. Right: Comparing estimation errors across different simulation setups.}
\label{second_order}
\end{figure}

To study the accuracy of GCR, we implemented a version of GCR which utilizes the second order approximation as detailed before. We then compared the first- and second-order estimates of heritability on our initial set of simulated data. The resulting second order estimates were almost identical to the results obtained by the first order approximation (figure \ref{second_order}

\subsection{Simulations with fixed effects}

To test that our method still generates accurate estimates of heritability even in the presence of fixed effects, we used the following simulation scenario: In addition to generating the genotypes as described earlier, for each simulated individual we generated a 0-1 ''sex" variable with probability $0.5$ for each value. We fixed the heritability to $0.5$ and the proportion of cases in the study to $0.3$, and defined the risk of individuals with $\text{sex}=0$ to $0.005$. The risk of individuals with $\text{sex}=1$ was either $0.005,0.01$ or $0.02$, i.e. the relative risk (RR) was $1,2$ or $4$. The thresholds were adjusted to depend on the sex accordingly. Phenotypes were generated after accounting for the fixed effects by way of changing the threshold as described earlier, and individuals were selected for the study based only on their phenotypes (i.e. independently of their sex). 

\begin{figure}[H]
\includegraphics[scale=0.7]{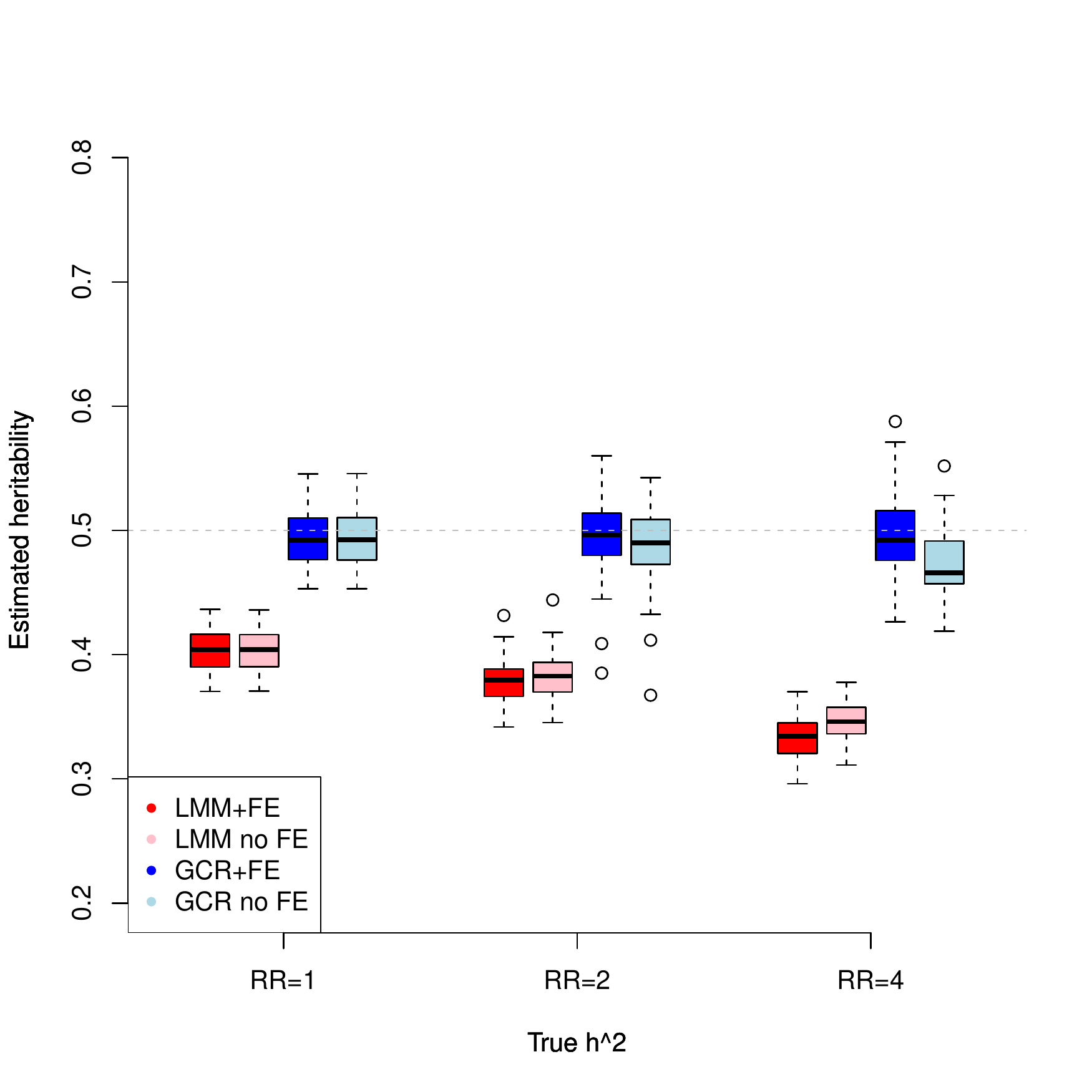}
\caption{Estimation of heritability with and without accounting for fixed effects using LMM and GCR. The true underlying heritability is $0.5$ and the only method which is consistently accurate is GCR with fixed effects.}
\label{Flo:fixed_effects}
\end{figure}

We simulated $100$ sets of genotypes/phenotypes/sex for each of the three possible RR values. We then estimated the heritability using LMM and GCR, each time with and without including the fixed effect in the analysis. The results are displayed in figure \ref{Flo:fixed_effects}. 
Our simulations demonstrate that GCR correctly accounts for fixed effects while LMM estimates are biased with and without accounting for fixed effects.

\section{Inference }

Standard inference for linear regression is not applicable in our
case due to the breaking of several of the basic assumptions used in
linear regression inference. For example, the errors are non-normal
and not independent. We therefore use the jackknife \citep{efron1994introduction} to estimate
the variance of our estimates. Since our method is three to four orders
of magnitude faster than LMM estimation, applying such a procedure
is not considerably time-demanding. 
More specifically, we use the jackknife to estimate the standard deviation of the estimate, and use this standard deviation to construct a confidence interval (CI). At each jackknife iteration we remove one individual from the data, and remove all relevant entries to the genetic correlation matrix. We constructed 95\% confidence intervals for our simulations and, these CIs covered the true value of the heritability in 98\% of the simulations, indicating the jackknife is conservative in this case, and the true standard error of the estimates is usually smaller than the jackknife estimates. 

\section{Heuristic correction of LMM estimates }

While we have demonstrated that our method can be used to yield accurate
estimates of heritability, it might also be of interest to correct
existing estimates of heritability which were obtained using LMM
and so might be biased. 

In their seminal work Dempster and Lerner \citep{dempster1950heritability} derive the relationship
between additive heritability on the observed scale and heritability
on the liability scale under the liability threshold model. However,
D\&L's work is done under a no-ascertainment assumption. Lee et
al. \citep{lee2011estimating} extend their result to ascertained case-control studies,
and obtain the following relationship:

\[
h_{o}^{2}=\varphi(t)^{2}\frac{P(1-P)}{K^{2}(1-K^{)2}}\frac{\sigma_{g}^{4}}{\sigma_{g_{cc}^{2}}}=\varphi(t)^{2}\frac{P(1-P)}{K^{2}(1-K^{)2}}\frac{h_{l}^{4}}{\sigma_{g_{cc}^{2}}}.\]

Note that the relationship depends on the ratio of variances, so that
the relationship is not linear. The variance of the genetic effects
under ascertainment is given by:

\[
\sigma_{g_{cc}}^{2}=\sigma_{g}^{2}\Big[1+\sigma_{g}^{2}\varphi(t)\frac{(P-K)}{K(1-K)}\Big]\Big[t-\varphi(t)\frac{(P-K)}{K(1-K)}\Big].\]

We note a different analytical expression of $\sigma_{g_{cc}}^{2}$
is given in Lee et al. We validated our derivation of this
expression numerically. A script demonstrating the correctness of
our derivation can be found on our webpage. 

Despite the fact that this is a relationship between parameters, and
we are interested in correcting estimates, we applied this correction
to the estimates obtained in our simulations. Our reasoning was twofold
- first, we hoped that the biases introduced by the estimation and
the biases introduced by using the observed scale would be interchangeable,
and thus correcting for the latter would expose the former, making
its analysis easier. Second, from \cref{Flo:REML_obs,Flo:REML_cor,Flo:reg_cor_box,Flo:low_asc,Flo:no_asc}, it seems that
LMM works well when heritability is low, regardless of $K$ and $P$, and also that LMM works well when ascertainment is low. We speculated that this is due to $\sigma_{g_{cc}}^{2}$ being close
to $\sigma_{g}^{2}$ in such scenarios. 

\begin{figure}[H]
\includegraphics[scale=0.5]{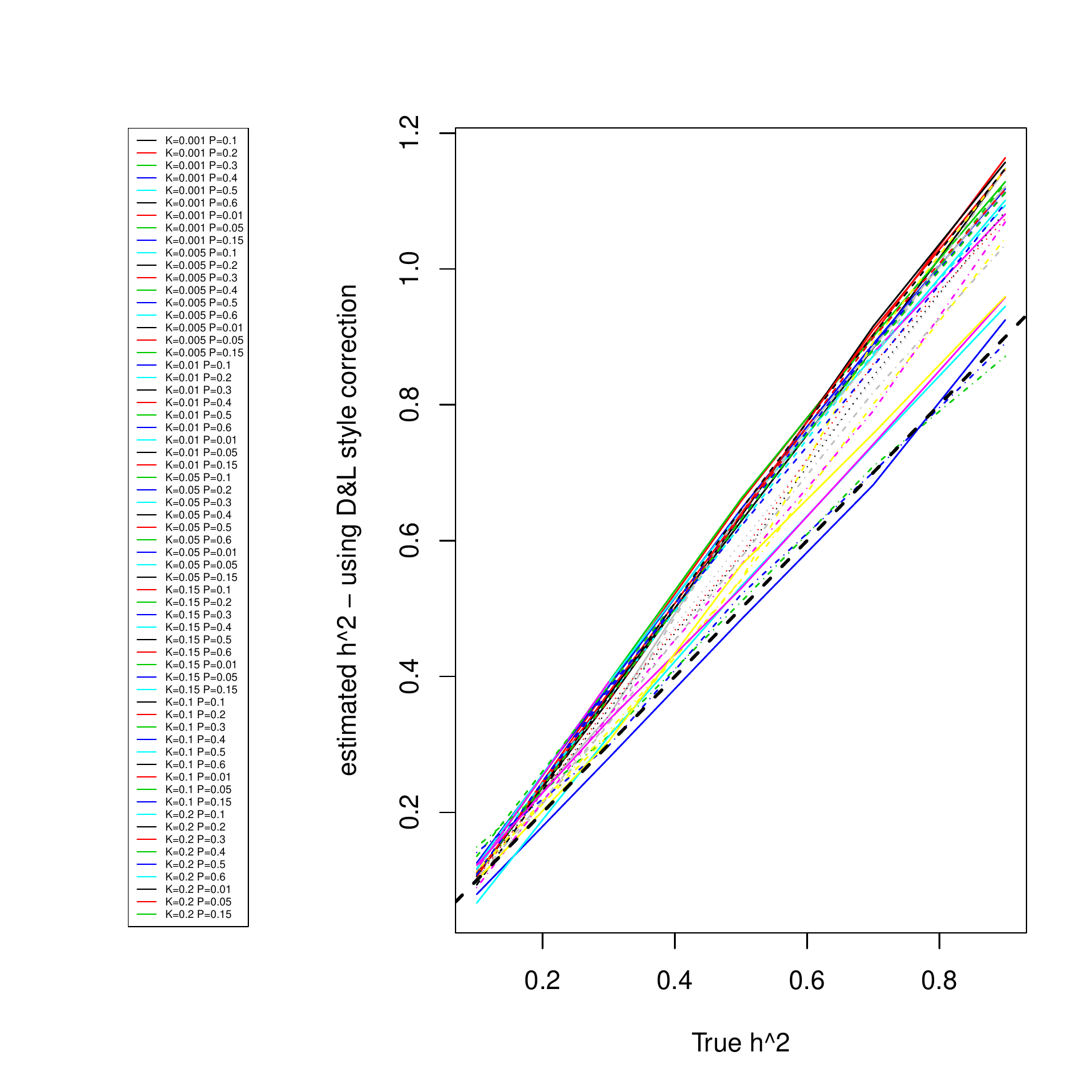}

\caption{Correcting LMM estimates using the generalization of Dempster\&Lerner
derived by Lee et al. produces linear dependency of the result on
the true underlying heritability, but the slope is not $1$, except
for the case of non-ascertained simulated studies. }
\label{Flo:DL_correction}
\end{figure}

The results of applying this correction to the simulations is shown
in figure \ref{Flo:DL_correction}. As can be seen from the figure,
the resulting estimates are still biased, except for the case of non-ascertained
studies, but the dependency of the corrected estimates on the true
underlying heritability seems linear. 

we speculated that the transformation indeed corrects for the effect of estimating
under the observed scale (rather than under the liability scale),
but do not correct for the additional effects introduced by the LMM
estimation. 

The resulting relationships between the corrected estimates and the
true underlying heritabilities seemed linear, but with a slope which
differs from $1$, we speculated that the difference was due to the
normality assumptions made by LMM, which are less accurate when ascertainment
is strong. We therefore attempted to model the slope as a function
of $K$ and $P$. Since for the no-ascertained case $K=P$ and the
slope is $1$, we speculated that the slope depends on $\frac{K}{P}$,
and that this size captures the degree of ascertainment. We therefore
estimated the slope for each $(K,P)$ tuple and explored the dependence
of the slope on the ratio $\frac{K}{P}.$

\begin{figure}[H]
\includegraphics[scale=0.5]{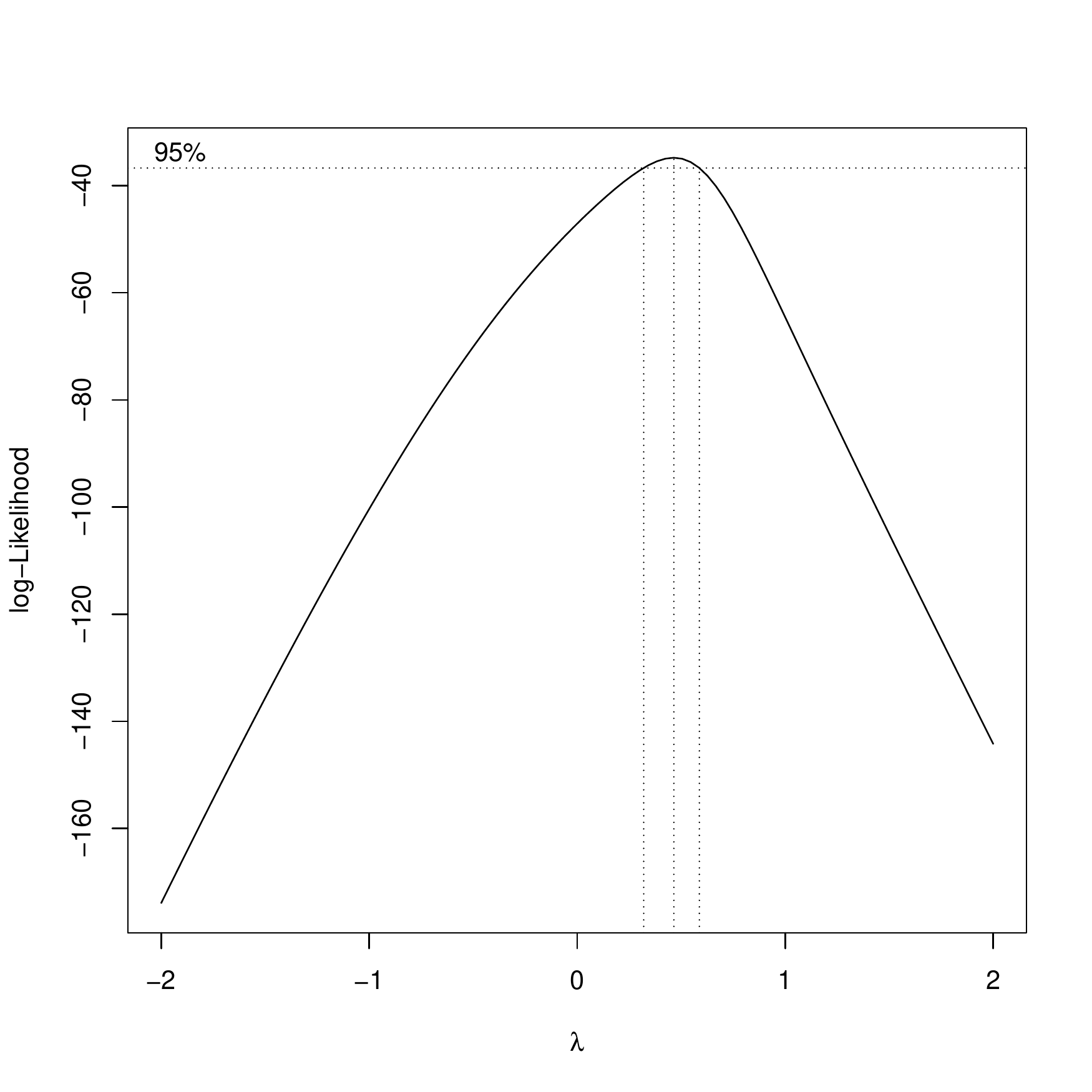}

\caption{A Box-Cox plot using $\frac{K}{P}$ as the dependent variable and
the slope as the covariate indicates that a square-root transformation
is a good fit. }
\label{Flo:boxcox}
\end{figure}

Using a Box-Cox plot \citep{box1964analysis} (figure \ref{Flo:boxcox}) indicated that $\sqrt{\frac{K}{P}}$
is the best transformation of the $\frac{K}{P}$ variable. 

We then used linear regression to estimate the relationship:

\[
\text{slope}(K,P)=1.3-0.3\sqrt{\frac{K}{P}}\]

\begin{figure}[H]
\includegraphics[scale=0.5]{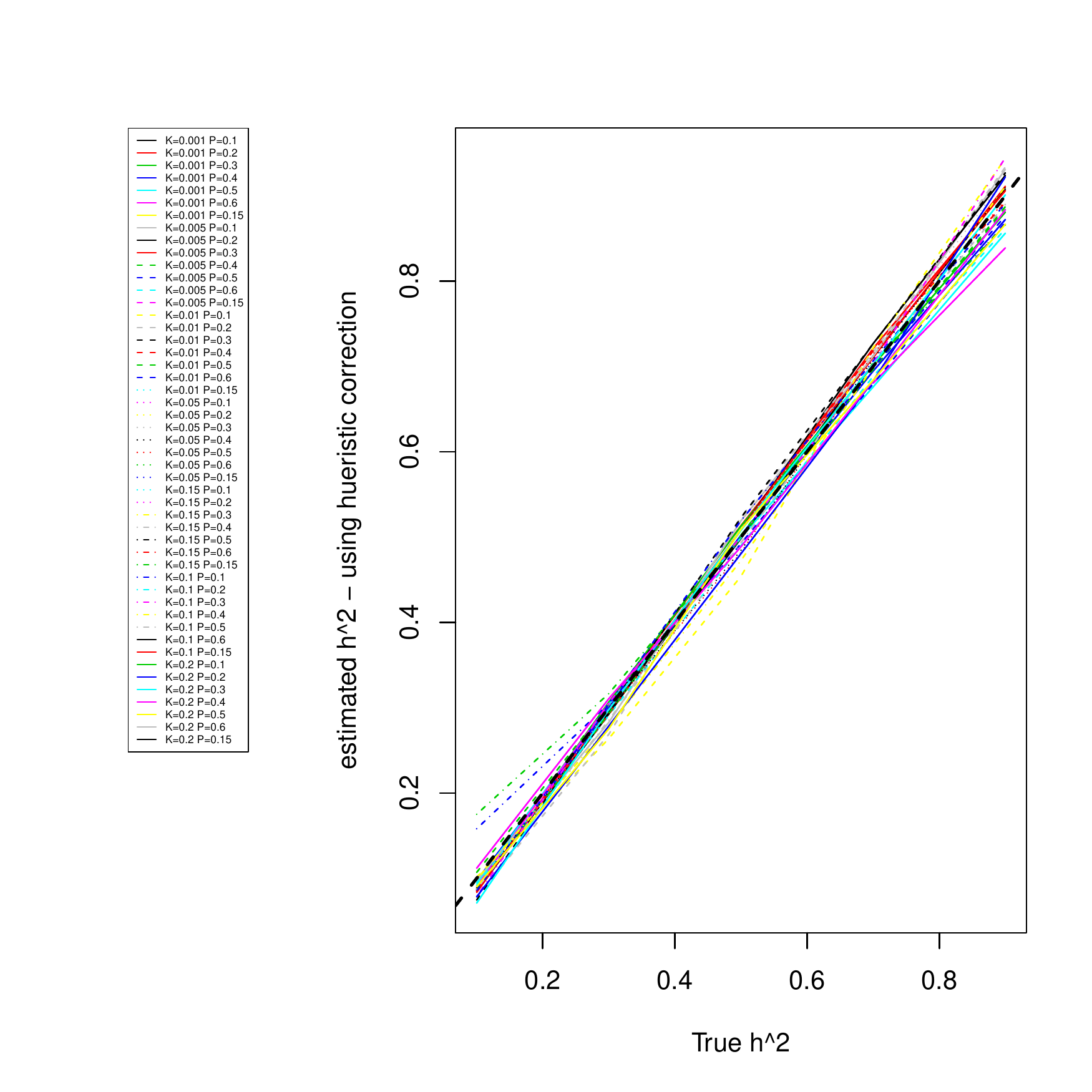}

\caption{Applying our final heuristic correction to LMM estimates. The dashed
line gives $y=x$.}
\label{Flo:final_huer}
\end{figure}

\begin{figure}[H]
\includegraphics[scale=0.5]{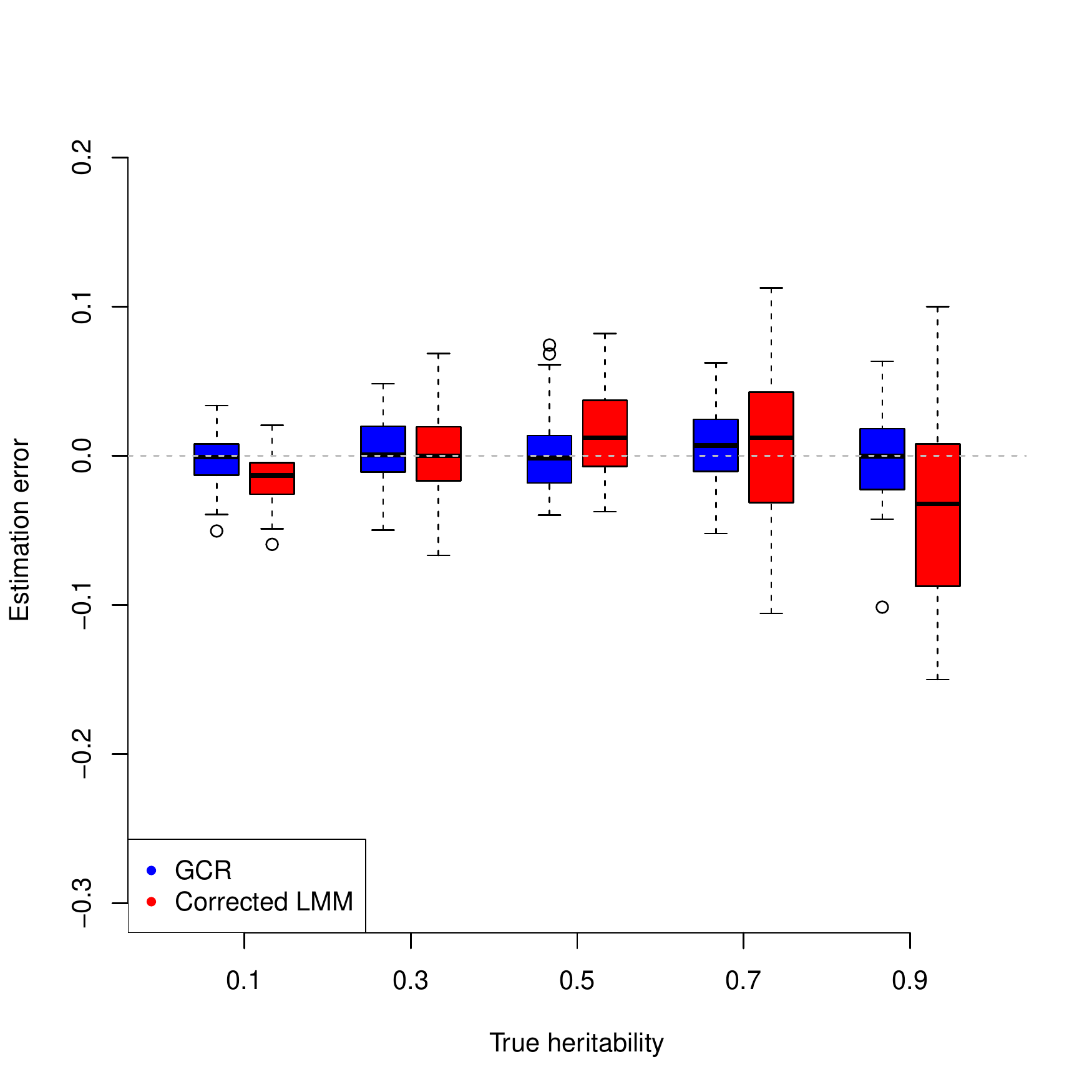}

\caption{Comparing estimation errors of GCR and corrected LMM.}
\label{Flo:final_huer_box}
\end{figure}

We concluded by applying this correction on top of the previous correction
(i.e. dividing the estimates by the slope). The results of this correction
are given in figure \ref{Flo:final_huer}. While the results are not
as well behaved as when using GCR, it seems that applying
these two corrections one on top of the other reduces the bias considerably
for all simulated ascertainment scenarios. While after correcting, the results are unbiased, in terms of variance, GCR still outperforms LMM (figure \ref{Flo:final_huer_box})

\subsection{Applying the correction to published data}

So far, we have derived our heuristic correction using simulations
wherein the true underlying heritability is known. However, when attempting
to correct published estimates, we only know $K,P$ and $\hat{\sigma_{g}^{2}}_{LMM}$.
We define our corrected estimate $\hat{\sigma_{g}^{2}}^{*}$ to be
the value for which \[
\mathbb{E}[\hat{\sigma_{g}^{2}}_{LMM};\sigma_{g}^{2}=\hat{\sigma_{g}^{2}}^{*}]=\hat{\sigma_{g}^{2}}_{LMM},\]

i.e. the value of underlying heritability for which the expected value
of the estimator is the observed estimate, where the expectation is
calculated using our heuristic correction. 

Confidence intervals are derived by applying the same procedure to
the top and bottom limits of a $95\%$ confidence interval based on
the published standard deviation of the estimate. 

\subsection{Testing the correction on WTCCC data}

\begin{figure}[H]
\includegraphics[scale=0.5]{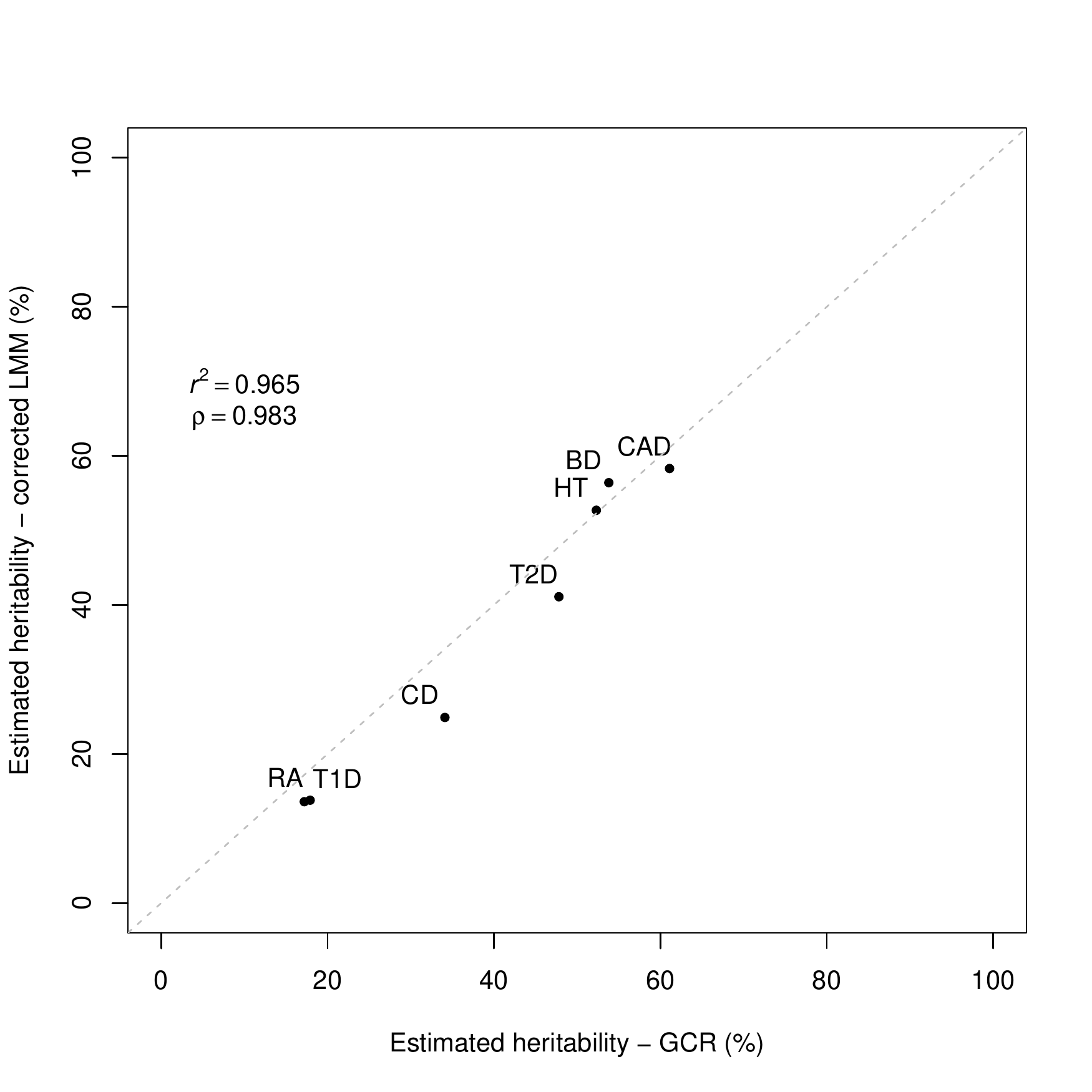}

\caption{Estimates for all 7 WTCCC phenotypes obtained by GCR, as compared
to a the estimates resulting from applying our heuristic correction
to LMM estimates. The dashed line gives $y=x$.}
\label{Flo:WTCCC_correction}
\end{figure}

To validate our suggested heuristic correction, and to make sure it
generalizes beyond our simulations, we computed LMM estimates of heritability
for all 7 WTCCC case-control studies, and applied the correction to
the resulting estimate. We also applied our method to the data and
compared the result to the corrected estimates. The results are highly
correlated ($\rho=0.97$) and can be seen in figure \ref{Flo:WTCCC_correction}.

\subsection{Applying our correction to other published results}
We applied our correction to several published case-control studies \citep{lee2011estimating,lee2012estimating,lee2013estimation,do2011web}. The corrected heritability estimates are given in table \ref{corrected_table}.

\begin{table}[H]
\begin{tabular}{|c|c|c|c|c|}
\hline 
Phenotype & K (\%) & P (\%) & LMM (\%)& Corrected LMM (\%)\tabularnewline
\hline
\hline 
CD & 0.1 & 39.2 & 22 [16.1-27.9] & 28.5 [17.7-44]\tabularnewline
\hline
 BD & 0.5 & 36.9 & 38 [30.2-45.8] & 58 [38.6-86.7]\tabularnewline
\hline
 T1D & 0.5 & 40.4 & 13 [5.2-20.8] & 12.3 [4.4-22.5]\tabularnewline
\hline
 SCH & 1 & 42.7 & 23 [21-25] & 24.4 [21.7-27.2]\tabularnewline
\hline
 ED & 8 & 31.1 & 26 [18.2-33.8] & 25.1 [17-33.9]\tabularnewline
\hline
 AD & 2 & 46.1 & 24 [18.1-29.9] & 24.4 [17.3-32.4]\tabularnewline
\hline
 MS & 0.1 & 45.1 & 30 [24.1-35.9] & 52.6 [34-83.5]\tabularnewline
\hline
 PAR & 1 & 10.4 & 27 [22.7-31.3] & 25.3 [20.8-30]\tabularnewline
\hline
\end{tabular}
\caption{Published and corrected LMM heritability estimates for several studies \citep{lee2011estimating,lee2012estimating,lee2013estimation,do2011web}}
\label{corrected_table}
\end{table}



\section{Accounting for imperfect linkage-disequilibrium}
Yang et al. \cite{yang2010common} suggest that the genetic correlation matrix, estimated based on the genotyped data, is in fact not the true underlying correlation matrix, as the latter should be computed only using causal SNPs. Since causal SNPs are unknown, and often not genotyped, the estimated genetic correlation is a noisy and biased estimate of the true genetic correlation. To correct for this effect, Yang et al. suggest using a different correlation matrix 
\[G^{*}=\beta(G-I)+I\]
where $G$ is the estimated genetic correlation matrix, and $\beta$ is a constant correcting the bias between the true and estimated genetic correlations. Using extensive simulations using real data and realistic assumptions they estimate $\beta$ at roughly $0.875$. It is easy to show that using $G^{*}$ in LMM estimation yields an estimate of heritability which is $\frac{1}{\beta}\hat{h}^2$, where $\hat{h}^2$ is the heritability estimate.
In the case of GCR, we regress the product of normalized phenotypes $Z_{ij}$ on $G_{ij}$. Hence, plugging in $G^{*}_{ij}=\beta G_{ij}$ as the covariate has the exact same effect of increasing the estimated heritability by a factor of $\frac{1}{\beta}$.

\section{Quality Control}

Following Lee et al. (2011) we applied a stringent quality control
(QC) process on the WTCCC data to avoid detecting spurious heritability
due to genotyping differences between cases and controls or between
the different control groups. We removed SNPs with MAF>5\%, with missing
rate >1\% and SNPs which displayed a significantly different missing
rate between cases and controls (p-value<0.05). We also removed SNPs
which deviated from Hardy-Weinberg (HW) equilibrium in the control groups
(p-value <0.05). Additionally we removed SNPs which displayed a significant
difference in frequency between the two control groups. Sex chromosomes
were excluded from the analysis. We removed all the individuals appearing
in the WTCCC exclusion lists. These include duplicate samples, first
or second degree relatives, individuals which are not of European
descent and other reasons. In addition we removed individuals with missing
rate >1\% and all individual pairs with an estimated relationship
>0.05 based on the correlation matrix. The last step is done to ensure individuals in the study are not closely related.

\section{No evidence of population structure}

Lee et al. \cite{lee2011estimating} include several top principal component vectors as fixed effects in an attempt make sure heritability estimates are not inflated due to population structure in the WTCCC samples. We suggest that this procedure is not required for two reasons.
First, we note that \cite{WTCCC} perform a thorough investigation of population structure in the WTCCC samples, and conclude that once individuals of different ancestry are removed from the samples, the effect of population structure is very minute. 

Second, \cite{patterson2006population} present a straightforward statistical test to decide how many principal components should be included as covariates in an association study. However, their method assumes that the individuals are randomly sampled from the population. In our case, cases are over-sampled in the study, and so LD regions hosting a causative variant might be mistaken for population structure. One simple way to overcome this problem is to dilute the number of SNPs, making sure each LD block is represented by a small number of SNPs. It is then unlikely that the differences between the case and control groups due to caustive variants would be captured by a top ranking principal component.

We applied the method of \cite{patterson2006population} do the GWAS data using every $k$'th SNP for increasing values of $k$. If there's a real population structure, we expect the number of significant principal components to be non-zero asymptotically. However, this was not the case, in agreement with \cite{WTCCC} (data not shown).

We note that including principal components which tag such regions hosting causative variants would result in an underestimation of the heritability, as such a vector would be highly correlated with the phenotype. It is therefore highly undesired to include principal components which do not tag actual population structure.

\bibliographystyle{plain}
\bibliography{main}